\documentclass[twocolumn]{aastex701}

\usepackage{subcaption}
\usepackage{placeins}

\begin{document}

\title{Systematic Monitoring of Extreme X-ray Variability from Weak-line Quasars}

\author[orcid=0009-0001-1311-1952]{Madison Reich}
\affiliation{Department of Astronomy and Astrophysics, 525 Davey Lab, The Pennsylvania State University, University Park, PA 16802, USA}
\affiliation{Institute for Gravitation and the Cosmos, The Pennsylvania State University, University Park, PA 16802, USA}
\email[show]{mjr7306@psu.edu}  

\author[orcid=0000-0002-0167-2453]{W. N. Brandt}
\affiliation{Department of Astronomy and Astrophysics, 525 Davey Lab, The Pennsylvania State University, University Park, PA 16802, USA}
\affiliation{Institute for Gravitation and the Cosmos, The Pennsylvania State University, University Park, PA 16802, USA}
\affiliation{Department of Physics, 104 Davey Laboratory, The Pennsylvania State University, University Park, PA 16802, USA}
\email[show]{wnb3@psu.edu}

\author[orcid=0000-0002-9036-0063]{Bin Luo}
\affiliation{School of Astronomy and Space Science, Nanjing University, Nanjing, Jiangsu 210093, China}
\affiliation{Key Laboratory of Modern Astronomy and Astrophysics (Nanjing University), Ministry of Education, Nanjing 210093, China}
\email{bluo@nju.edu.cn}

\author[orcid=0000-0002-8577-2717]{Qingling Ni}
\affiliation{Max-Planck-Institut f\"ur extraterrestrische Physik (MPE), Gie\ss enbachstra\ss e 1, D-85748 Garching bei M\"unchen, Germany}
\email{qni@mpe.mpg.de}

\author[0000-0002-7092-0326]{Richard M. Plotkin}
\affiliation{Department of Physics, University of Nevada, Reno, NV 89557, USA}
\affiliation{Nevada Center for Astrophysics, University of Nevada, Las Vegas, NV 89154, USA}
\email{rplotkin@unr.edu}

\author[orcid=0000-0003-4327-1460]{Ohad Shemmer}
\affiliation{Department of Physics, University of North Texas, Denton, TX 76203, USA}
\email{ohad.shemmer@unt.edu}

\author[orcid=0000-0003-0680-9305]{Fabio Vito}
\affiliation{INAF – Osservatorio di Astrofisica e Scienza dello Spazio, Via Gobetti 93/3, I-40129, Bologna, Italy}
\email{fvito.astro@gmail.com}

\author[0000-0001-9314-0552]{Weimin Yi}
\affiliation{Yunnan Observatories, Kunming, 650216, People's Republic of China}
\email{wmyi2012@gmail.com}

\begin{abstract}
We present the results of a multi-cycle \textit{Chandra} program to systematically monitor the X-ray variability of 10 weak-line quasars (WLQs) that previously had limited multi-epoch X-ray observations. Three new \textit{Chandra} 2.8 to 8.2 ks observations were obtained for each WLQ with $\mathrm{C}\,\text{\scriptsize IV}$ rest-frame equivalent widths (REWs) $\lesssim 10$ \AA, substantially improving the monitoring data quality of WLQs and our ability to characterize their long-term X-ray variability behavior. We observe recurrent extreme X-ray variability in the historically variable WLQ SDSS J1539+3954, with an X-ray flux rise of a factor of $\gtrsim 6$ between 2023 and 2024 ($\gtrsim 21$ relative to 2013). Another previously X-ray weak WLQ in the sample, SDSS J0825+1155, underwent a significant X-ray flux variation by a factor of $\gtrsim 14$ between 2019 and 2023. We find the fraction of WLQs exhibiting evidence of extreme X-ray variability to be $0.20^{+0.17}_{-0.07}$. In the context of the thick disk and outflow (TDO) model, the substantial fraction of WLQs displaying extreme X-ray variability may suggest that the variability is {more likely} driven by the intrinsic motion of the TDO wind rather than by changes in the height of the TDO disk. We performed a statistical comparison between the distribution of variability amplitudes of WLQs and general radio-quiet quasars. We find that these underlying distributions are statistically different, with WLQs having $\approx 6.8$ times higher odds of exhibiting an extreme X-ray variability event than the general radio-quiet quasar population.
\end{abstract}

\section{Introduction}
\subsection{Observations and Models for Weak-Line Quasars}
Weak-line quasars (WLQs) and their remarkable observed properties present a valuable opportunity to test theories of quasar accretion physics and structure.  WLQs are a subclass of luminous Type 1 quasars, with the blue UV/optical continua typical of quasars. However, WLQs lack the characteristic strong broad high-ionization emission lines expected of a Type 1 quasar under the AGN unification model. They instead have remarkably weak or absent broad-emission lines, with $\mathrm{C}\,\text{\scriptsize IV}$
 rest-frame equivalent widths (REWs) $\lesssim 10$ \AA \ and/or Ly$\alpha$+$\mathrm{N}\,\text{\scriptsize V}$ REW $\lesssim 15$ \AA, corresponding to $\gtrsim 3\sigma$ negative deviations from the REW distribution mean \citep{1999ApJ...526L..57F,2009ApJ...699..782D,2010ApJ...721..562P}. These weak high-ionization lines also tend to show very large blueshifts of \mbox{$1000$--$10000 \ \mathrm{km \ s^{-1}}$} (e.g., \citealt{2012ApJ...747...10W}). WLQs preferentially show high values of $\mathrm{C}\,\text{\scriptsize IV}\  ||$ Distance \citep{2023ApJ...950...97H}\footnote{{Due to the weak nature of the $\mathrm{C}\,\text{\scriptsize IV}$ emission line, measurements of $\mathrm{C}\,\text{\scriptsize IV}$ REW and blueshift can be subject to systematic uncertainties, resulting in larger measurement errors. Reliable $\mathrm{C}\,\text{\scriptsize IV}$ measurement methodologies are addressed in, e.g., \citet{2023ApJ...950...95M}, \citet{2023ApJ...950...96D}, and \citet{2023ApJ...950...97H}.}}, a parameter mapping the combination of the rest-frame $\mathrm{C}\,\text{\scriptsize IV}$ equivalent width and $\mathrm{C}\,\text{\scriptsize IV}$ blueshift in a two-dimensional $\mathrm{C}\,\text{\scriptsize IV}$ parameter space (e.g., \citealt{2020MNRAS.492.4553R, 2022ApJ...931..154R}). The $\mathrm{C}\,\text{\scriptsize IV}\  ||$ Distance  parameter is likely correlated with the accretion rate, $L/L_{\mathrm{Edd}}$ \citep{2020MNRAS.492.4553R, 2022ApJ...931..154R}, suggesting that WLQs accrete at high rates.
 
 The low-ionization broad-emission lines, e.g., $\mathrm{Mg}\,\text{\scriptsize II}$, $\mathrm{H\beta}$, and $\mathrm{H\alpha}$, of WLQs are not as strikingly weak as the high-ionization lines {(e.g., \citealt{2025ApJ...994..213C})}. For example, \citet{2024ApJ...972..191C} show the distribution of WLQ $\mathrm{H\beta}$ REWs to be consistent with that of a typical population of radio-quiet quasars (WLQ average $\mathrm{H\beta}$ REW \mbox{$\approx 52.9 \pm 2.0$ \AA} \ compared to $\mathrm{H\beta}$ REW $\approx 64.4 \pm 8.1$ \AA \ for typical radio-quiet quasars), and \citet{2015ApJ...805..123P} show a generally weaker WLQ $\mathrm{H\beta}$ REW $\approx 15\text{--}40$ \AA. The population of $\approx$400 Sloan Digital Sky Survey (SDSS)-selected WLQs are primarily radio-quiet quasars, and are not BL Lac-type objects (e.g., \citealt{2009ApJ...696..580S,2010ApJ...721..562P}). The {observed} IR-to-UV spectral energy distributions (SEDs) of WLQs are largely consistent with the general quasar population {within the typical $1\text{--}2\sigma$ scatter of individual quasars about the mean SED}, e.g., \citet{2009ApJ...699..782D}, \citet{2011ApJ...743..163L},   \citet{2015ApJ...805..122L}, \citet{2018MNRAS.480.5184N}.  

Investigations of {the X-ray properties of WLQs} have been illuminating for our understanding of the nature of WLQs, especially when considered alongside their optical/UV spectral properties (e.g., \citealt{2015ApJ...805..122L,2018ApJ...865...92M,2018MNRAS.480.5184N,2022MNRAS.511.5251N}). Radio-quiet, non-Broad Absorption Line (BAL) quasars have been shown to follow an empirical relationship between their UV luminosity, $ L_{2500 \ \text{\AA}}$, and $\alpha_{\mathrm{OX}}$, where $\alpha_{\mathrm{OX}}$ is the power-law slope between the rest-frame 2500~{\AA} and 2 keV luminosities. 
About half ($\approx$48\%) of the WLQ population is X-ray weak compared to the X-ray emission expected based on their UV emission strength, often by large factors of 20–70. Whereas, the other half of the WLQ population has the nominal amount of X-ray emission expected from a luminous quasar. The X-ray weak WLQs constitute a much larger fraction of the total WLQ population than the X-ray weak objects in the general population of radio-quiet quasars do, where X-ray weak sources are $\lesssim 3\%$ of the total population (e.g., \citealt{2020ApJ...900..141P}). 

Additionally, stacking analyses of X-ray weak WLQs have shown that, on average, X-ray weak WLQs have hard X-ray spectra, with an effective power-law photon index of $\langle \Gamma_{\mathrm{eff}} \rangle \approx 1.2\text{--}1.4$ \citep{2018MNRAS.480.5184N}. The hard spectra of X-ray weak WLQs are likely indicative of X-ray absorption via partial covering and/or Compton reflection. In line with this idea, studies of individual WLQs have revealed evidence of strong intrinsic X-ray absorption, e.g., the WLQ SDSS J1521+5202 investigated in \citet{2015ApJ...805..122L} and \citet{2024ApJ...974....2W}. In contrast, X-ray normal WLQs tend to have steep power-law spectra of $\langle \Gamma \rangle = 2.2 \pm 0.1$ \citep{2015ApJ...805..122L,2018ApJ...865...92M}, suggesting accretion at a high Eddington ratio (e.g., \citealt{2008ApJ...682...81S,2013MNRAS.433.2485B}).

Multi-epoch X-ray observations of WLQs have revealed a notable number of cases of large-amplitude X-ray variability, where the WLQ appears to transition between X-ray weak and X-ray normal states. \textit{XMM-Newton} observations of the WLQ PHL 1092 revealed a 2 keV flux-density drop of a factor $\sim 260$ on a timescale of years, while the UV emission remained nearly constant \citep{2012MNRAS.425.1718M}. PHL 1811, observed by \citet{2007ApJ...663..103L} as an intrinsically X-ray weak WLQ by a factor of $\approx 30\text{--}100$, underwent an X-ray flux brightening by a couple orders of magnitude, as observed by the Einstein Probe in 2024 \citep{2024ATel16763....1L,2026arXiv260609981L}. SDSS J1521+5202 displayed a rise of 0.5--2.0 keV flux up to a factor of $\approx 32$ between 2006 and 2023 \citep{2024ApJ...974....2W}. \citet{2022ApJ...930...53L} reported a 0.5--2.0 keV flux drop by a factor of $\approx 7.6$ over two rest-frame days in the WLQ SDSS J1350+2618.

The WLQ SDSS J153913.47+395423.4 (hereafter J1539+3954) was originally observed in an X-ray weak state in 2013, but its 0.5–2.0 keV flux rose by a factor of $\gtrsim 20$ between 2013 and 2019 \citep{2020ApJ...889L..37N}. A follow-up with \textit{Chandra} Director’s Discretionary Time in 2020 observed another extreme variability event with a drop in X-ray flux by a factor of $\approx$ 9, showing J1539+3954 to return to an X-ray weak state on the time scale of a few months \citep{2022MNRAS.511.5251N}. Contemporaneous rest-frame UV spectroscopic observations with the Hobby-Eberly Telescope (HET) show that the UV continuum and emission lines remain largely unchanged, despite the extreme variability behavior in the X-ray. {Disk reverberation mapping of J1539+3954 with data from the Zwicky Transient Facility (ZTF) has also revealed a $\sim$33-day (observed frame) time delay between the ZTF \textit{g}- and \textit{r}-band light curves \citep{2023ApJ...956..126M}. This interband time lag is not consistent with accretion-disk reprocessing of variable X-ray emission, which may suggest Compton-thick X-ray obscuration in J1539+3954, or the presence of additional mechanisms beyond X-ray reprocessing that could be driving the UV/optical variability \citep{2023ApJ...956..126M}.} Additionally, radio follow-up of J1539+3954 by \citet{2026ApJ...996...23C} after its observed large-amplitude brightening and dimming in the X-ray showed no significant variability in the radio flux during the period of large-amplitude X-ray variations, suggesting no direct connection between the X-ray variability and radio emission.

The X-ray variability amplitude for the population of typical luminous radio-quiet quasars generally does not exceed a factor of $\gtrsim 2\text{--}3$ (see \citealt{2020MNRAS.498.4033T} and discussion below). The number of WLQs with large-amplitude X-ray variability events appears to be large compared to the expectations of typical radio-quiet quasars (e.g., \citealt{2020MNRAS.498.4033T}). The number of WLQs with extreme X-ray variability may suggest that weak high-ionization emission lines are an indicator of extreme X-ray variability in luminous quasars. 

A shielding-gas model {was derived empirically} to explain the various multiwavelength properties of WLQs {(e.g., \citealt{2011ApJ...736...28W,2025ApJ...994..213C})}. In this model, there exists a body of shielding gas on small-scales ($\lesssim 30 R_S$) that blocks the ionizing EUV/X-ray photons from reaching the broad emission-line region (BELR), resulting in the observed weak UV broad emission lines. This shielding gas must also be optically thick to high-energy X-ray photons ($N_H \gg 10^{24} \ \mathrm{cm^{-2}}$) to explain the X-ray weak WLQ population via strong X-ray absorption along our line of sight intercepting this shielding gas. {\citet{2015ApJ...805..122L}, \citet{2018MNRAS.480.5184N}, and \citet{2022MNRAS.511.5251N} built upon this originally empirical model by suggesting that the source of the shielding gas is} a geometrically and optically thick inner accretion disk and outflow{, hereafter the thick disk and outflow (TDO) model, as} expected for quasars accreting at high Eddington ratios (e.g., \citealt{2014ApJ...796..106J, 2019ApJ...885..144J, 2014MNRAS.439..503S, 2014ApJ...797...65W}). The proposed accretion at a high Eddington ratio is {consistent} with observations of steep X-ray power-law spectra in X-ray normal WLQs (e.g., \citealt{2015ApJ...805..122L,2018ApJ...865...92M}).\footnote{{Single-epoch virial mass estimates of WLQs based on FWHM(H$\beta$) have been shown to underestimate the black-hole (BH) mass, and subsequently overestimate accretion rate (e.g., \citealt{2022MNRAS.515..491M,2020ApJ...897..117M,2018NatAs...2...63M}). Furthermore, in the super-Eddington regime, the virial assumption needed to derive single-epoch virial BH masses breaks down in the presence of large radiation pressure (e.g., \citealt{2025ApJ...995..171C}). In the empirically developed TDO model, we avoid any reliance upon single-epoch virial mass estimates by utilizing the steep intrinsic X-ray continuum as independent evidence for high Eddington rates in WLQs (see \citealt{2015ApJ...805..122L,2018ApJ...865...92M,2025ApJ...995..171C}).}}.

In the context of the TDO model, the extreme X-ray variability events likely originate from changes in the TDO that expose or obscure the central X-ray source. The TDO model is consistent with the observed spectral transitions between X-ray normal and X-ray weak states in WLQs. High levels of line-of-sight X-ray absorption and/or Compton reflection due to shielding gas will result in an X-ray weak state with a hard X-ray spectrum{, as soft X-rays are more easily photoelectrically absorbed than hard X-rays}. Conversely, a reduction in the X-ray absorption along the line of sight will lead to a softer spectral shape and a return to an X-ray normal state, as the shielding gas no longer obscures the observer’s line of sight. The exact cause of the changes in the structure of the TDO is unknown. However, further investigations into the fraction of WLQs that display such large-amplitude X-ray variability events will allow us to begin to constrain the cause of TDO variability, {as the fraction of WLQs exhibiting extreme X-ray variability can be used as a basic proxy for the covering fraction of the variable region of the TDO.} Potential explanations for the variability in the physical structure of the TDO include the following:

\begin{enumerate}
    \item If the extreme X-ray variability is caused by slight changes in the height of a rotating TDO moving across our line of sight, this would occur only in the objects where our line of sight skims the boundary, or ‘surface,’ of the TDO. The covering fraction of the region driving variability would be small, and as such, only a small fraction of WLQs would exhibit such extreme X-ray variability, though still a larger fraction than observed in typical quasars.

    \item The extreme X-ray variability could instead be driven by the intrinsic motion of the TDO wind, such as the movement of clumpy, optically-thick material across our line of sight. {The motion of clumpy winds in the TDO model would produce variable X-ray absorption along our line of sight, giving rise to the extreme X-ray variability behavior observed. The covering fraction of this region of variability in the TDO would be larger than in the boundary case above, and thus} a larger fraction of WLQs may display this behavior. 

    \item Lastly, we consider the possibility that the X-ray variability could be driven by a larger, global reconfiguration of the TDO wind. In this scenario, the fraction of WLQs displaying extreme X-ray variability would be dependent on the frequency of such large-scale reconfigurations. 
\end{enumerate}

Alternative explanations for the observed weak broad emission lines of WLQs have been proposed. One such explanation proposed an ‘anemic’ BELR, where the BELR has an unusually low gas-content (e.g., \citealt{2010ApJ...722L.152S}). Another suggested that the high-ionization lines are weak because the BELR is exposed to an intrinsically soft ionizing continuum (e.g., \citealt{2004ApJ...611..125L}). These models, however, are challenged by {the observed X-ray properties of WLQs, as they cannot account for the large fraction of X-ray weak WLQs, the X-ray spectral properties of WLQs, or the observed X-ray variability.}

\subsection{A Systematic Study of Weak-Line Quasar X-ray Variability}
Prior to this work, studies of the X-ray variability of WLQs were largely limited to serendipitous discoveries of large-amplitude variability, as opposed to a targeted observing program monitoring a sample of WLQs for this behavior. Additionally, the number of available multi-epoch X-ray observations was limited, and was too small to place systematic constraints on the long-term X-ray variability behavior of WLQs. Due to the serendipity and sparseness of the available data, the previously observed WLQs of interest had been selected following different sample selection criteria to serve a wide range of scientific applications. For these reasons, a study based on the limited X-ray observations of WLQs could not be conducted effectively. 

Despite the limitations in monitoring data quality, remarkable X-ray variability results have been found, regarding both the large-amplitude X-ray variations observed and the frequency at which such events occur. These results have motivated a more careful study of the long-term X-ray variability behavior of WLQs. We have therefore performed a three-cycle \textit{Chandra} observing campaign for a set of 10 WLQs previously having limited (1--3) multi-epoch observations. This observation program has provided us with 4--6 sensitive \textit{Chandra} epochs for each WLQ, marking a large improvement in the monitoring data quality and in our ability to characterize the long-term X-ray behavior of WLQs. We additionally utilize contemporaneous optical light curves from ZTF (e.g., \citealt{2019PASP..131a8003M}) to monitor for any optical variability coincident with X-ray variations. 

\subsection{Paper Overview \& Definitions}

This paper is organized as follows. We describe the selection criteria of the WLQ sample and the \textit{Chandra} observations in Section 2. In Section 3, we outline the X-ray photometric analysis and optical light curve construction, and derive the X-ray and optical properties per observation. Section 4 presents a summary of the variability behavior of the WLQ sample, and describes a statistical comparison between the variability amplitudes of WLQs and the general population of radio-quiet quasars. We summarize the results and discuss future work in Section 5. {In Appendix A, we list the \textit{Chandra} observations used in this work and their derived X-ray and optical properties. In Appendix B, we provide the ZTF \textit{g}/\textit{r}/\textit{i}-band light curves for each WLQ in the sample. Appendix C lists the results of the X-ray variability amplitude analysis of the WLQ sample.}

In this paper, we use J2000 coordinates, and a cosmology with $H_0 = 67.4 \ \mathrm{km \ s^{-1} \ Mpc^{-1}}$, $\Omega_M = 0.315$, and $\Omega_{\Lambda} = 0.685$ (\!\!\citealt{2020A&A...641A...6P}).

\section{Sample Selection \& \textit{Chandra} Observations} 
For the systematic monitoring of the X-ray variability behavior of WLQs, we targeted 10 WLQs from a representative sample of 32 SDSS-selected, radio-quiet ($R < 10$), non-BAL WLQs with $\mathrm{C}\,\text{\scriptsize IV}$
 REW $\lesssim 10$ \AA \ 
 from \citet{2018MNRAS.480.5184N}. The \citet{2018MNRAS.480.5184N} sample was constructed from the \citet{2011ApJS..194...45S} SDSS Data Release 7 (DR7) quasar catalog. We required our sample to be optically bright ($m_i \leq 18.1$) and thus luminous ($M_i \approx -27 \ \mathrm{to}  -28$), as quasars that are optically bright also tend to be X-ray bright. 

Our sample consists of eight WLQs classified as X-ray weak ($\Delta \alpha_{\mathrm{OX}} < -0.2$)\footnote{$\Delta \alpha_{\mathrm{OX}} = \alpha_{\mathrm{OX}} - \alpha_{\mathrm{OX}}(L_{2500 \ \text{\AA}})$. A complete derivation of $\Delta \alpha_{\mathrm{OX}}$ is presented in detail in Section 3.3.2.} and two classified as X-ray normal ($\Delta \alpha_{\mathrm{OX}} \approx 0$) prior to this monitoring campaign, allowing us to systematically monitor for both large-amplitude brightening and dimming in X-ray flux. The X-ray weak WLQs targeted are the eight most X-ray weak (corresponding to $\Delta \alpha_{\mathrm{OX}}$ values of \mbox{$\lesssim-0.33 \  \text{to} -0.56$)} of the optically bright WLQs from the representative WLQ sample in \citet{2018MNRAS.480.5184N}. The historically variable WLQ J1539+3954, most recently observed to be in an X-ray weak state at the start of this monitoring campaign, falls within this selection of X-ray weak WLQs \citep{2022MNRAS.511.5251N}. The two targeted X-ray normal WLQs are the optically brightest sources in the full representative WLQ sample \citep{2018MNRAS.480.5184N}.

Our systematic monitoring campaign took place from 2022 January 28 to 2024 November 03 spanning \textit{Chandra} Cycles 23--25, with each WLQ observed once per cycle. The \textit{Chandra} observations of this program were taken using the Advanced CCD Imaging Spectrometer spectroscopic array (ACIS-S) \citep{2003SPIE.4851...28G} in VFAINT mode. To remain sensitive to X-ray flux brightening by factors of $\gtrsim 3\text{--}8$ and X-ray flux dimming by factors of $\gtrsim 10$, the exposure times of each epoch ranged from 2.8 to 8.2 ks.

We additionally utilize 1--3 archival \textit{Chandra} observations for each WLQ in our sample from the preceding WLQ observations in \textit{Chandra} Cycles 11, 12, 14, 17, and 21 \citep{2011ApJ...736...28W, 2012ApJ...747...10W, 2015ApJ...805..122L, 2018MNRAS.480.5184N, 2022MNRAS.511.5251N}.  With the archival \textit{Chandra} epochs, we are able to sample the long-term behavior of X-ray variability on timescales of $\approx 7\text{--}15$ observed-frame years with 4--6 epochs per WLQ. The sample has a median of 5 epochs per WLQ. We list all (archival and new) \textit{Chandra} X-ray observations of the WLQs in our sample {in machine readable format, as described in Appendix~\ref{sec:appendixA}.} 

\section{Observations \& Data Analysis}

\subsection{\textit{Chandra} X-ray Photometric Analysis} 
Following the same methodology as in \citet{2022MNRAS.511.5251N}, we utilized \textit{Chandra}’s data analysis software, CIAO \citep{2006SPIE.6270E..1VF}, to process the archival and new \textit{Chandra} observations. We ran the \mbox{CHANDRA\_REPRO} script to create reprocessed observations, following the recommended steps in the CIAO data processing threads. We then extracted the full-band, soft-band, and hard-band images (0.5--8.0 keV, 0.5--2.0 keV, and 2.0--8.0 keV, respectively) with the standard set of ASCA grades. We used WAVDETECT on the full-band image to determine the source position for each observation. In the event of a non-detection, where WAVDETECT returns no source within 1 arcsecond of the SDSS position, we adopt the SDSS position as the source position. 

We defined a 2\arcsec \ radius circular aperture for the source region and a background annulus with an inner radius of 10\arcsec \ and an outer radius of 40\arcsec. We manually checked that no sources fell within our defined background apertures. We extracted the background light curve for each image and removed background flares at a 3-sigma level using the DEFLARE script to run an iterative sigma-clipping algorithm. This de-flaring procedure resulted in minimal data loss.

We performed aperture photometry to extract the aperture-corrected net counts in the full, soft and hard bands. We then computed the binomial no-source probability, $P_B$, {the probability that the putative detected source is not real given the local background level of the observation,} as a measure of the significance of the source detection in each band \citep{2007ApJS..169..353B,2011ApJS..195...10X, 2015ApJ...805..122L}. {$P_B$ is defined as}
\begin{equation}
P_B(X \geq S) = \sum^N_{X = S} \frac{N!}{X!(N-X)!}p^X(1-p)^{N-X}
\end{equation}
{where $S$ is the number of counts in the source region, $B$ is the number of counts in the background region, and $N = S + B$. In this expression, $p = 1/(1 + BACKSCAL)$, where $BACKSCAL$ is the ratio of the area of the background region to the area of the source region.} If, in a given band, $P_B < 0.01$, we considered the source detected in that band, and calculated the $1 \sigma$ uncertainty on the net counts \citep{1986ApJ...303..336G}. The 90\% confidence upper limits are provided by \citet{1991ApJ...374..344K} in the event of a non-detection, as determined by $P_B > 0.01$. From our detection threshold of $P_B < 0.01$, we expect to have $< 1$ false detection in our sample. 

We computed the band ratio, defined as the ratio of the hard-band to soft-band counts, and its uncertainty using fastHR\footnote{\url{https://github.com/fanzou99/FastHR}} \citep{2023ApJ...950..136Z}. If a source was significantly detected in both the hard and soft bands, we estimated the $1 \sigma$ uncertainty on the band ratio. We estimated the 90\% confidence level upper limit on the band ratio when the source was only detected in the soft band. 

\subsection{X-ray Results \& X-ray-to-optical Properties}
\subsubsection{X-ray Properties}
For each WLQ, we adopt values of Galactic column density from the HI 4 Pi Survey map of neutral hydrogen (\!\!\citealt{2016A&A...594A.116H}). The derived $N_{H,\mathrm{Gal}}$ values for each WLQ are listed in Table~\ref{tab:wlq_properties}. We then used MODELFLUX to derive the effective power-law photon index, $\Gamma_{\mathrm{eff}}$, and its uncertainty from the previously derived band ratio, assuming a power-law model with Galactic absorption. For observations where we are unable to constrain the effective photon index usefully from the data due to a non-detection of the source or a large uncertainty on the derived effective photon index ($\left<1 \sigma \right> > 0.5$), we adopted an effective photon index based on previous work. We adopt an effective photon index of $\Gamma_{\mathrm{eff}} = 1.4$ for observations where the targeted WLQ is X-ray weak ($\Delta \alpha_{\mathrm{OX}} < -0.2$), {as the stacking analysis of 30 X-ray weak WLQs in \citet{2015ApJ...805..122L} shows hard X-ray spectra with $\left<\Gamma_{\mathrm{eff}}\right> = 1.4$.} Similarly, we assume an effective photon index prior of $\Gamma_{\mathrm{eff}} = 2.2$ for the observations where the targeted WLQ is X-ray normal ($\Delta  \alpha_{\mathrm{OX}} > -0.2$) following from the results of \citet{2015ApJ...805..122L} and \citet{2018ApJ...865...92M}. 

\begin{table}
\centering
\renewcommand{\arraystretch}{1.25}
\setlength{\tabcolsep}{5pt}   \footnotesize
\caption{Properties of the WLQ Sample \label{tab:wlq_properties}}

\begin{tabular}{ccccc}
\hline
Object name & 
$z$ & 
$N_{H,\mathrm{Gal}}$ & 
$f_{2500\,\text{\AA}}$ & 
log$L_{2500\,\text{\AA}}$ \\
(1) & (2) & (3) & (4) & (5) \\
\hline
082508.75+115536.3 & 1.998 & 4.09 & 4.62 & 31.67 \\
094533.99+100950.0 & 1.671 & 2.38 & 3.42 & 31.40 \\
095023.19+024651.7 & 1.882 & 3.73 & 1.82 & 31.22 \\
110409.96+434507.0 & 1.804 & 0.88 & 1.75 & 31.17 \\
122855.90+341436.9 & 2.147 & 1.55 & 3.29 & 31.58 \\
140701.59+190417.9 & 2.004 & 2.65 & 1.78 & 31.26 \\
150921.68+030452.7
 & 1.808 & 3.56 & 5.35 & 31.66 \\
153913.47+395423.4& 1.935 & 1.71 & 3.86 & 31.56 \\
163810.07+115103.9 & 1.983 & 3.97 & 2.52 & 31.40 \\
215954.46\text{--}002150.1 & 1.965 & 4.28 & 7.94 & 31.89 \\
\hline
\end{tabular}

\tablecomments{
(1) J2000 object SDSS name. (2) Object redshift. (3) Column density of Galactic neutral hydrogen in units of $10^{20} \ \mathrm{cm^{-2}}$ \mbox{(\!\!\citealt{2016A&A...594A.116H})}. (4) Rest-frame 2500 \AA\ flux density in units of $10^{-27} \ \mathrm{erg \ cm^{-2} \ s^{-1} \ Hz^{-1}}$ from the \citet{2011ApJS..194...45S} SDSS DR7 quasar catalog. (5) Logarithm of the 2500 \AA\ luminosity in units of $\mathrm{erg \ s^{-1} \ Hz^{-1}}$.}
\end{table}

\begin{figure*}[htp!]{
    \centering

    \begin{subfigure}[b]{0.3\textwidth}
        \includegraphics[width=1.1\textwidth]{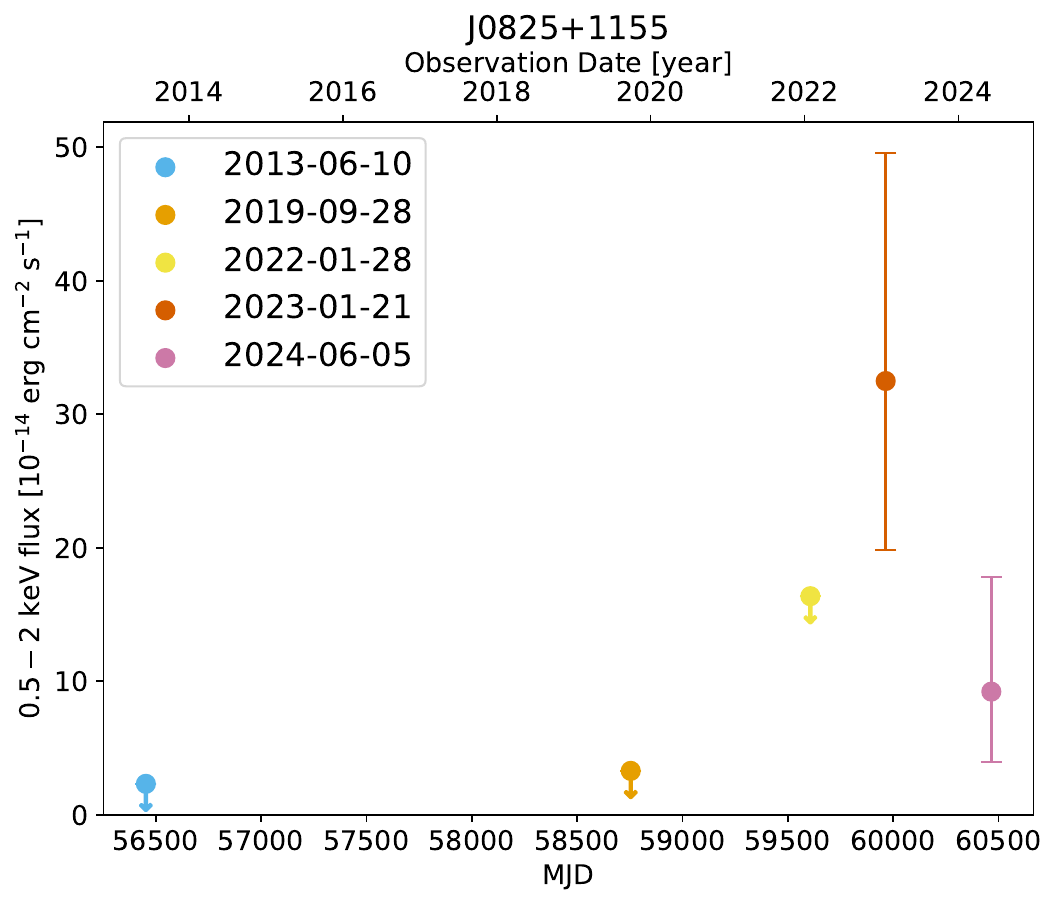}
    \end{subfigure}
    \hfill
    \begin{subfigure}[b]{0.3\textwidth}
        \includegraphics[width=1.1\textwidth]{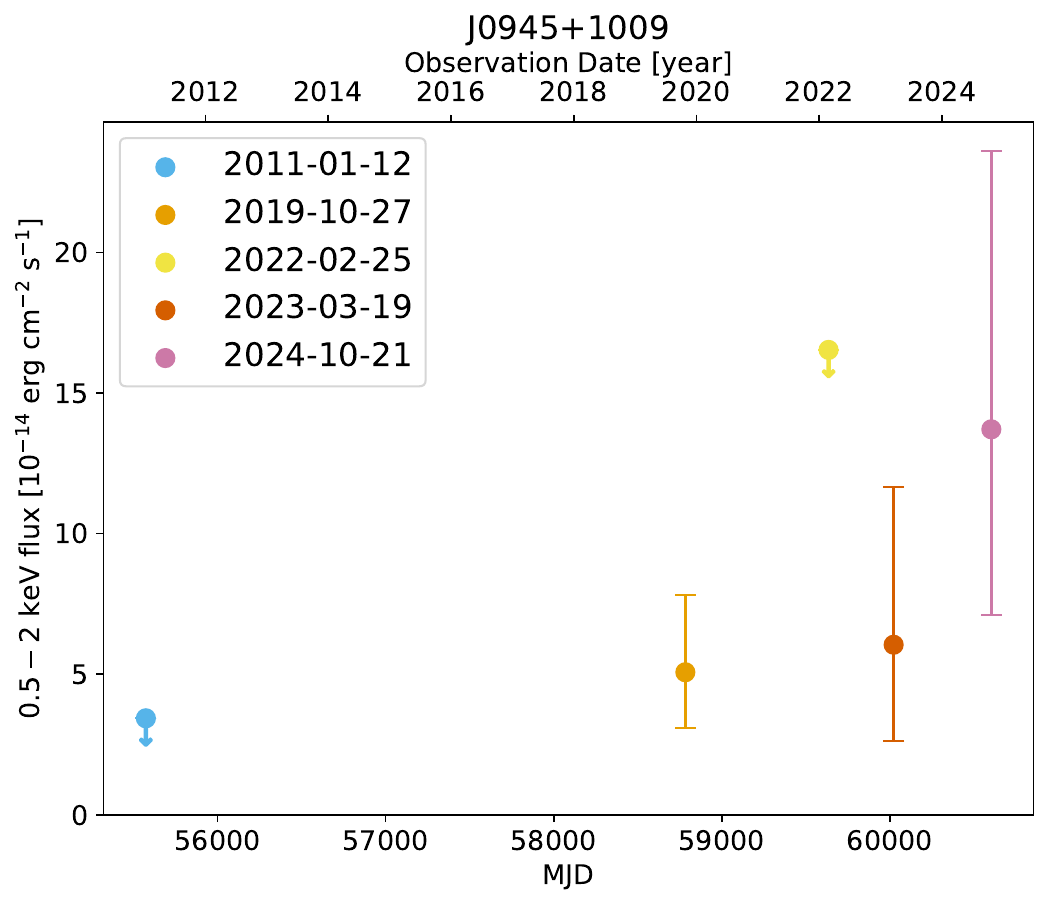}
    \end{subfigure}
    \hfill
    \begin{subfigure}[b]{0.3\textwidth}
        \includegraphics[width=1.1\textwidth]{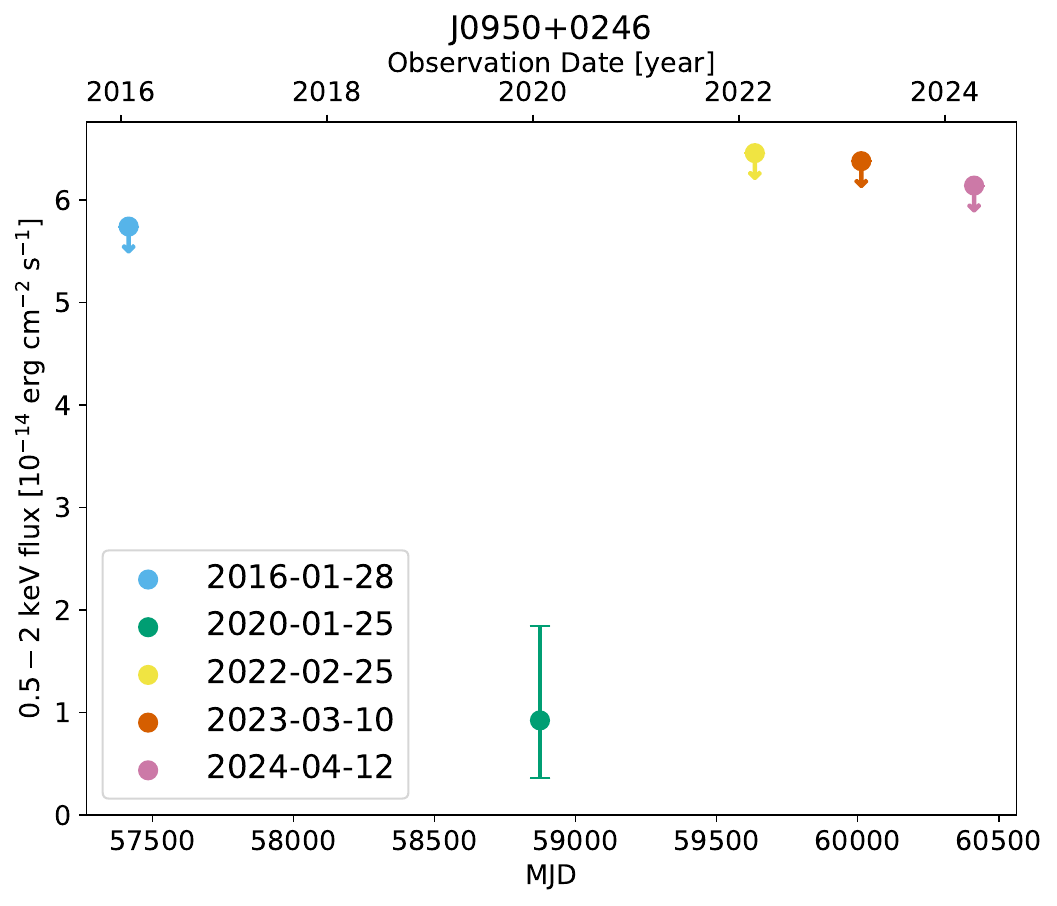}
    \end{subfigure}
    \hfill
    \begin{subfigure}[b]{0.3\textwidth}
        \includegraphics[width=1.1\textwidth]{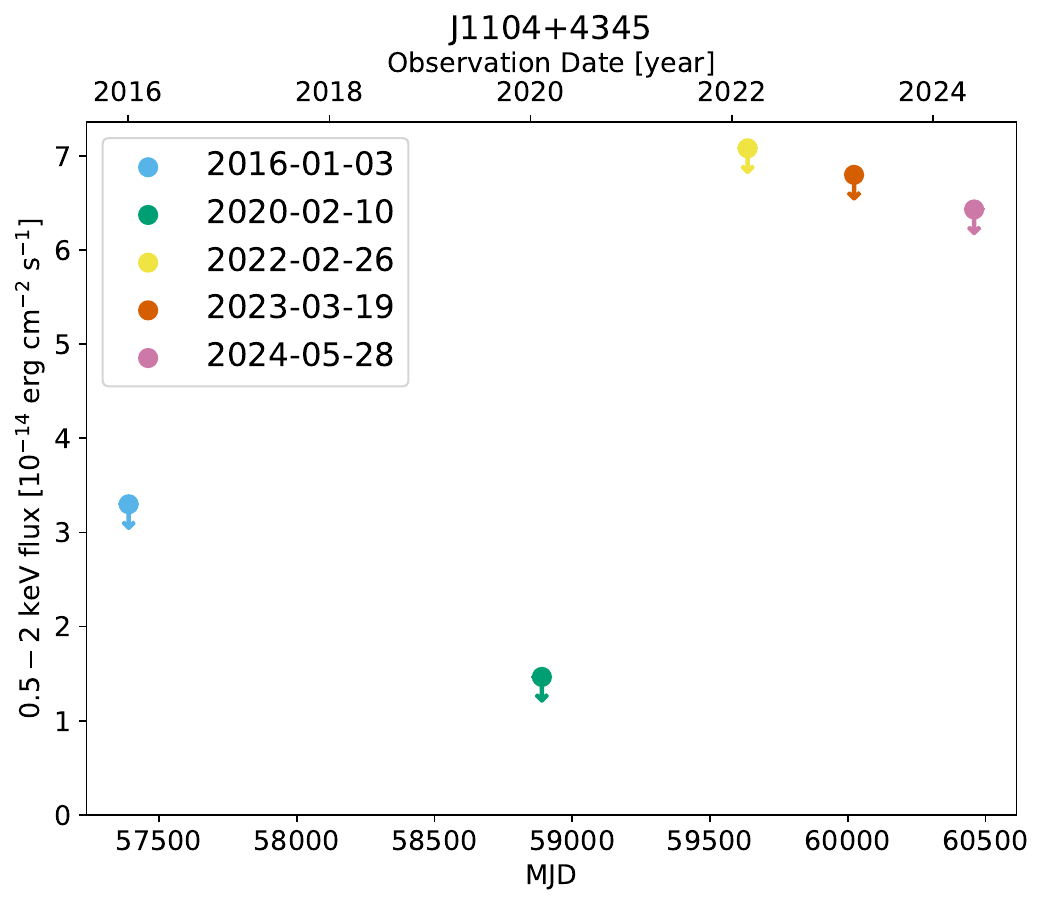}
    \end{subfigure}
    \hfill
    \begin{subfigure}[b]{0.3\textwidth}
        \includegraphics[width=1.1\textwidth]{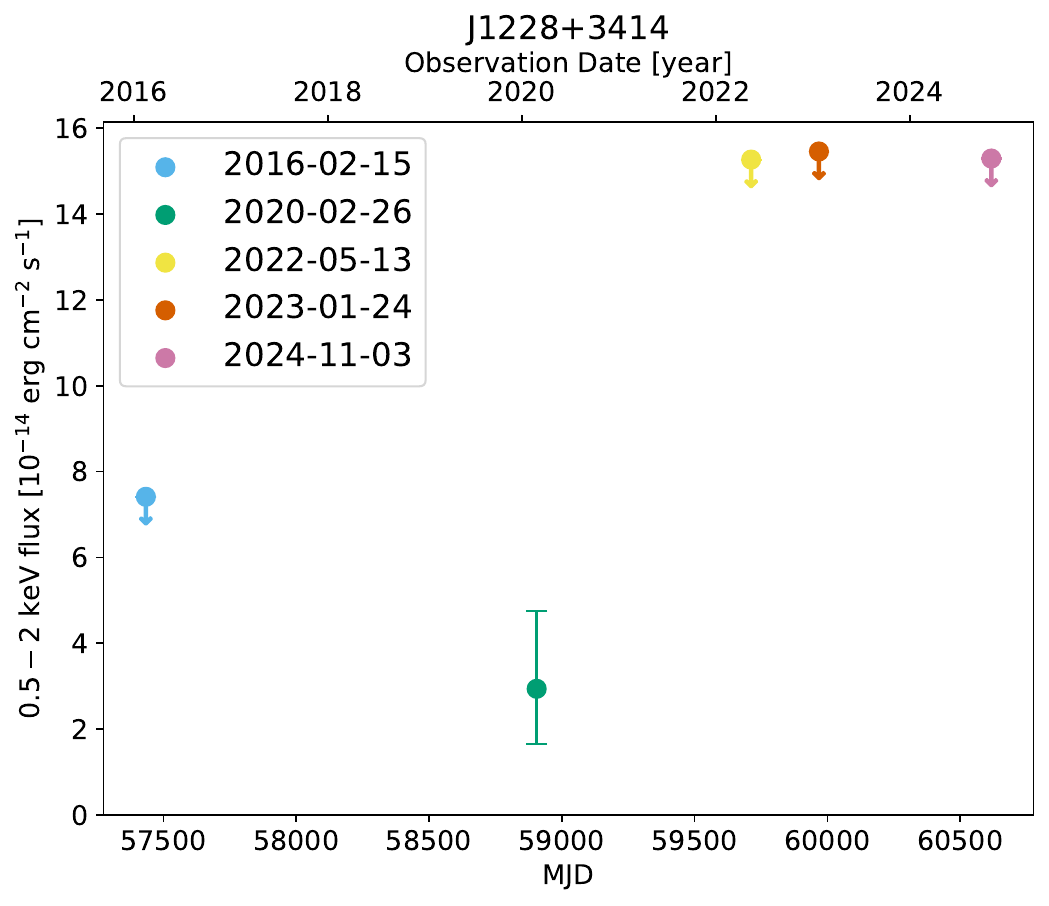}
    \end{subfigure}
    \hfill
    \begin{subfigure}[b]{0.3\textwidth}
        \includegraphics[width=1.1\textwidth]{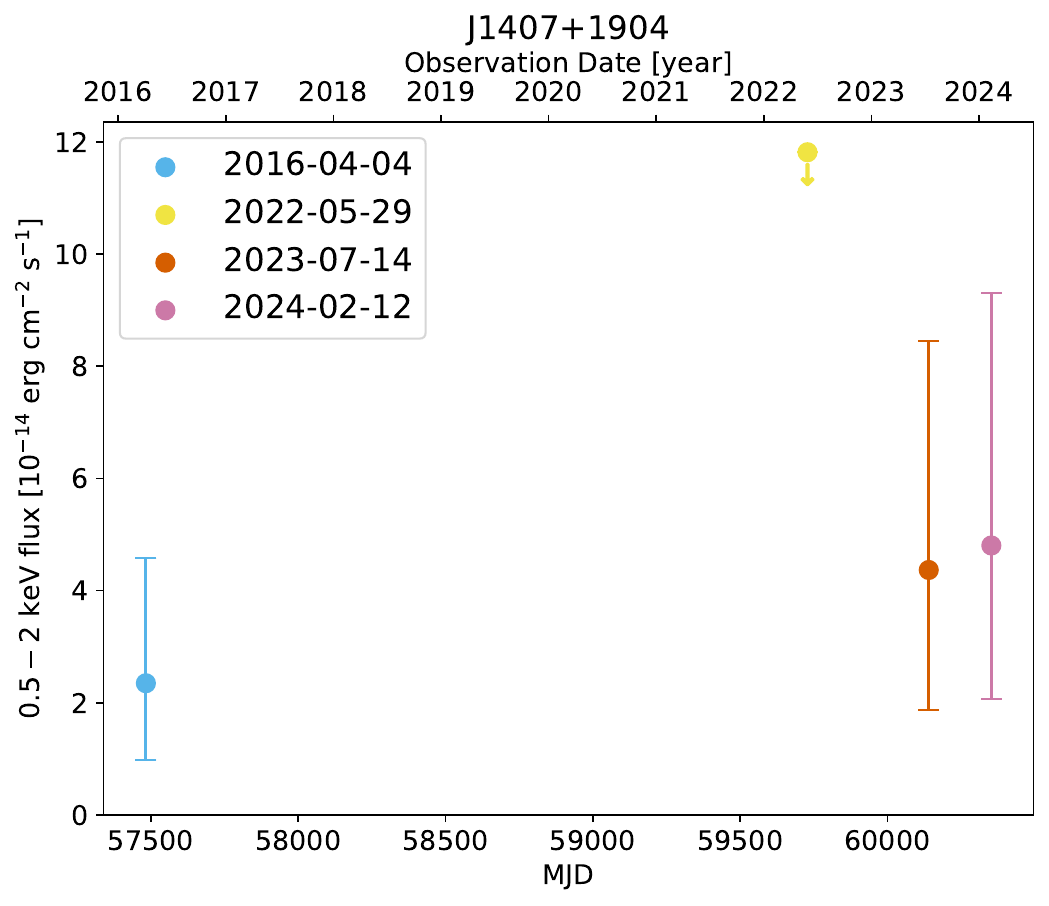}
    \end{subfigure}
    \hfill
    \begin{subfigure}[b]{0.3\textwidth}
        \includegraphics[width=1.1\textwidth]{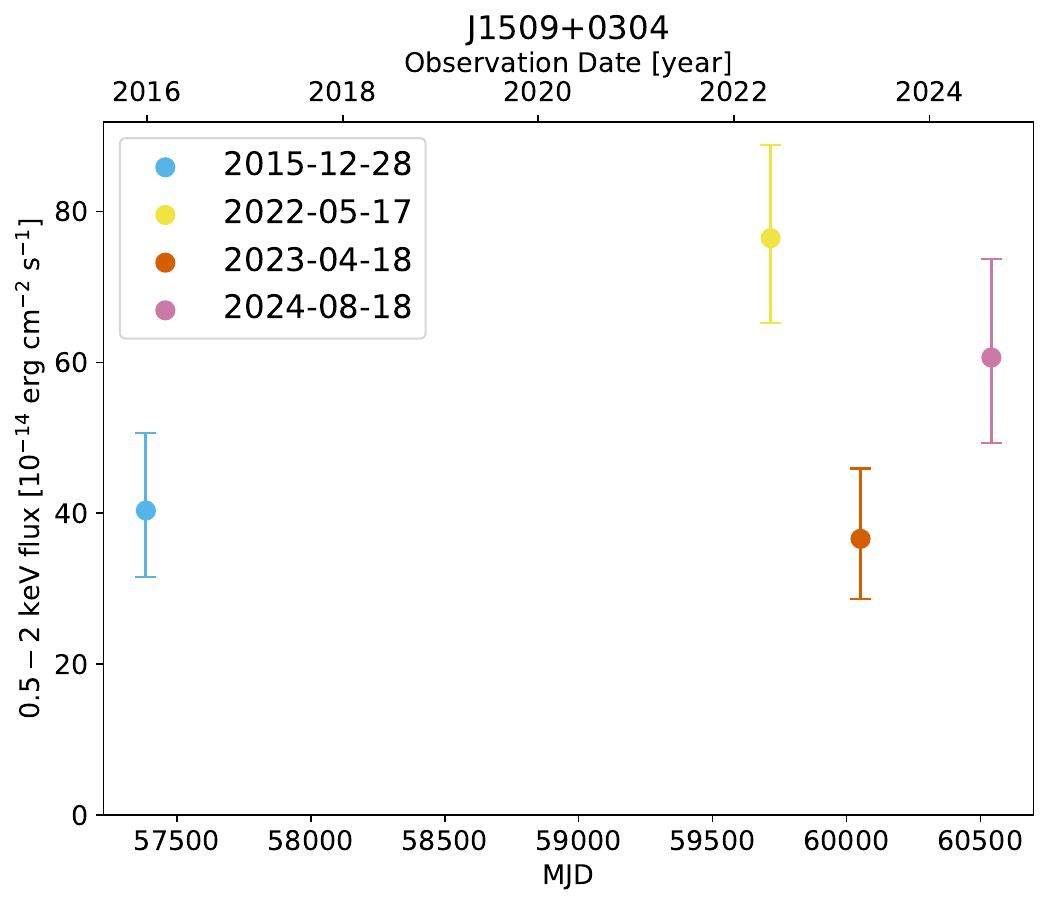}
    \end{subfigure}
    \hfill
    \begin{subfigure}[b]{0.3\textwidth}
        \includegraphics[width=1.1\textwidth]{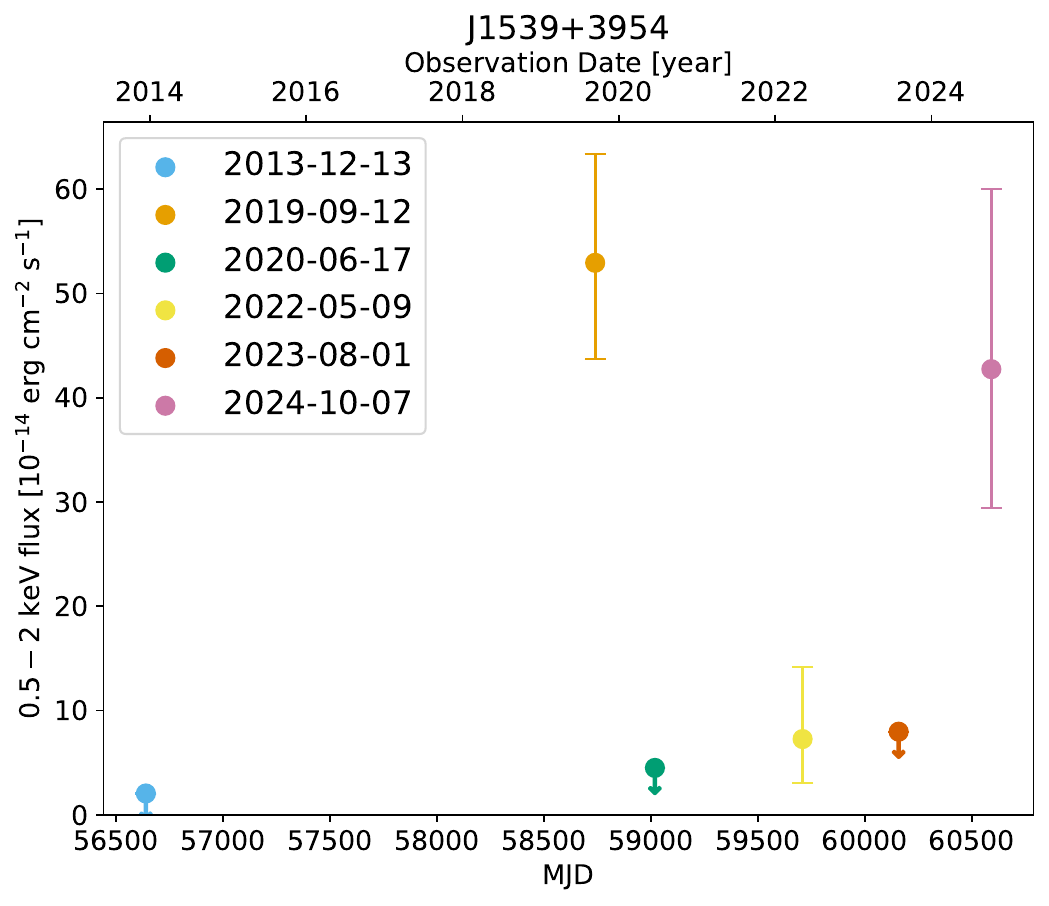}
    \end{subfigure}
    \hfill
    \begin{subfigure}[b]{0.3\textwidth}
        \includegraphics[width=1.1\textwidth]{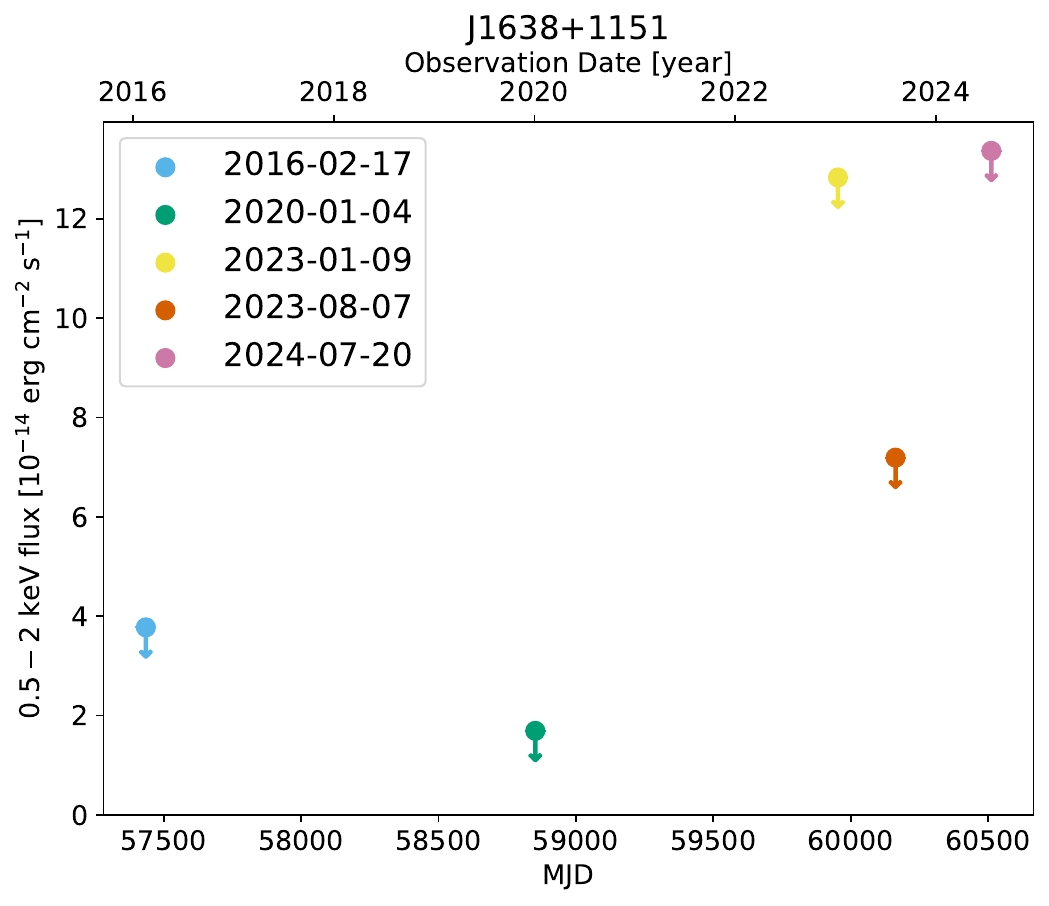}
    \end{subfigure}
    \hfill
    \begin{subfigure}[b]{0.3\textwidth}
        \includegraphics[width=1.1\textwidth]{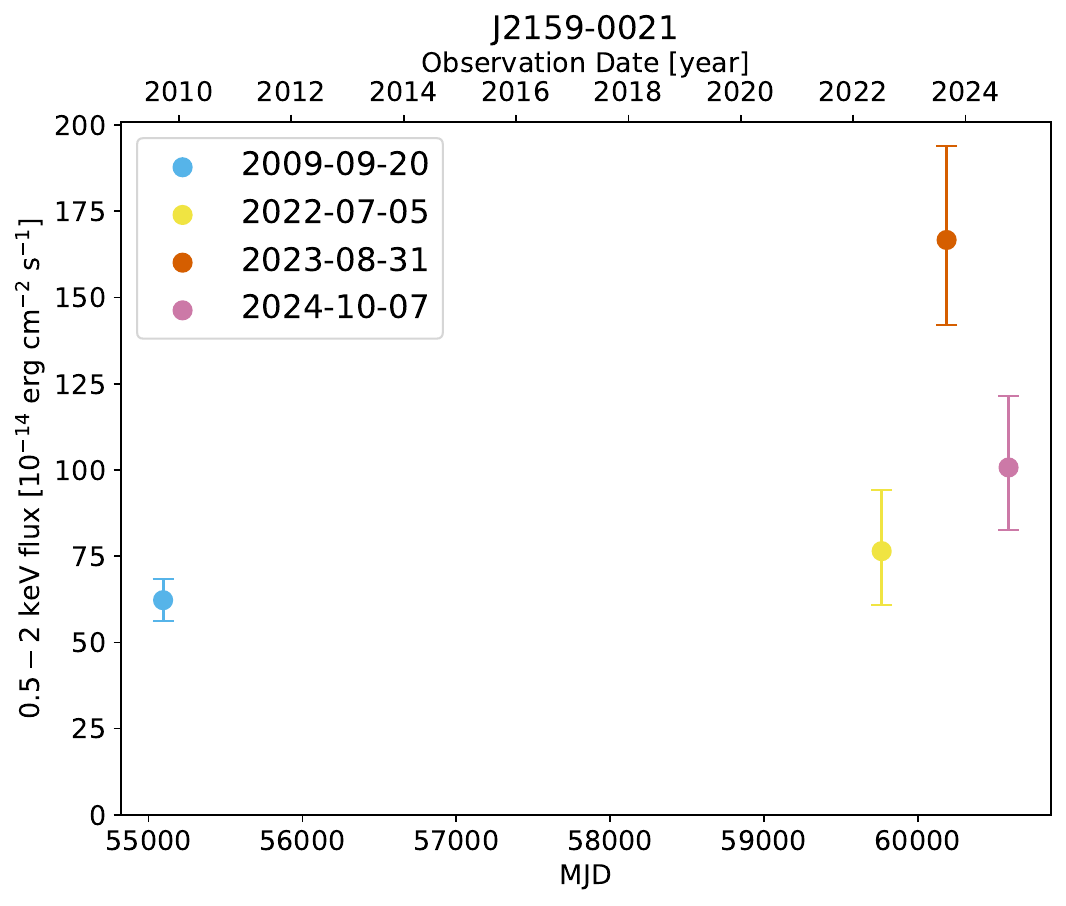}
    
    \end{subfigure}
    
    \caption{
        Long-term 0.5--2 keV X-ray light curves for the sample of WLQs. Solid points indicate significant source detections with 1$\sigma$ error bars. {Large 1$\sigma$ error bars are driven by low count statistics.} Points with downward arrows represent 90\% confidence level upper limits on the flux from instances of non-detections. {The colors represent the different X-ray epochs. Light blue points indicate the first archival epoch for each WLQ (spanning \textit{Chandra} Cycles 11–17), while orange and green points correspond to observations obtained in 2019 and 2020 (\textit{Chandra} Cycle 21), respectively. Yellow, red, and pink points represent observations from \textit{Chandra} Cycles 23, 24, and 25, respectively.}
    }
    \label{fig:xray_lc}
    }
\end{figure*}

We used SRCFLUX and the derived $\Gamma_{\mathrm{eff}}$ to calculate the Galactic-absorption corrected 0.5--2.0 keV flux, $F_X$, from the soft-band net count rate. We report the unabsorbed 0.5--2.0 keV flux and its $1\sigma$ uncertainty, or the 90\% confidence level upper limit if the source is undetected in the soft-band. We show the long-term \mbox{0.5--2.0 keV} light curve of each WLQ in Figure~\ref{fig:xray_lc}. From the unabsorbed soft-band flux and $\Gamma_{\mathrm{eff}}$, we then calculated the rest-frame 2 keV flux density, $f_{2 \ \mathrm{keV}}$. 

\subsubsection{X-ray-to-optical Properties}
We calculated the $\alpha_{\mathrm{OX}}$ value (upper limit) from the derived rest-frame 2 keV flux density value (upper limit)  and the rest-frame 2500 \AA \ flux density, $f_{2500 \ \text{\AA}}$. {For archival \textit{Chandra} X-ray epochs predating the start of the ZTF survey in 2018, we use $f_{2500 \ \text{\AA}}$ values from the \citet{2011ApJS..194...45S} SDSS quasar catalog. The $f_{2500 \ \text{\AA}}$ and log$L_{2500 \ \text{\AA}}$ values from \citet{2011ApJS..194...45S} are given in Table~\ref{tab:wlq_properties}. We derive contemporaneous values of the rest-frame 2500 \AA \ flux density for each \textit{Chandra} X-ray epoch from available ZTF optical light curves (see Section 3.3). We assume a typical quasar optical power-law spectral index of $-0.5$ and estimate $f_{2500 \ \text{\AA}}$ by extrapolating from the flux density measured at the central wavelength of the ZTF band closest to rest-frame 2500 \AA \ ($r$- or $i$-band for the redshift range of our WLQ sample). We visually inspected the SDSS spectra of each WLQ to confirm that an optical power-law spectral index of $-0.5$ is a good approximation. For WLQ SDSS 163810.07+115103.9 at $z = 1.983$, rest-frame 2500 \AA \ falls within the ZTF $i$-band, which contains no photometric data. Instead, we extrapolate from the central wavelength of the ZTF $r$-band to estimate $f_{2500 \ \text{\AA}}$ for this WLQ. We note that the extrapolation from the central wavelength of the ZTF band to rest-frame 2500 \AA \ is a small effect (flux change $\lesssim10\%$). } 

The X-ray to optical power-law slope, $\alpha_{\mathrm{OX}}$, is defined as $0.3838  \mathrm{log}(L_{2 \ \mathrm{keV}}/ L_{2500 \ \text{\AA}})$, and can be equivalently calculated as $\alpha_{\mathrm{OX}} = 0.3838   \mathrm{log}(f_{2 \ \mathrm{keV}}/  f_{2500 \ \text{\AA}})$. {We note that even for the range of $10\text{--}30  \% $ variability at rest-frame 2500 \AA, the resultant range in derived $\alpha_{\mathrm{OX}}$ is small ($0.02\text{--}0.05$) and does not significantly impact our results.} Using the empirical \ $ L_{2500 \ \text{\AA}}\text{--} \alpha_{\mathrm{OX}}$ relation from \citet{2007ApJ...665.1004J}, we calculated the expected value of $\alpha_{\mathrm{OX}}$, $ \alpha_{\mathrm{OX}}(L_{2500 \ \text{\AA}})$. We then calculated the difference between the derived $\alpha_{\mathrm{OX}}$ and the expected value, $ \alpha_{\mathrm{OX}}(L_{2500 \ \text{\AA}})$, $\Delta \alpha_{\mathrm{OX}} = \alpha_{\mathrm{OX}} - \alpha_{\mathrm{OX}}(L_{2500 \ \text{\AA}})$, to assess the deviation from the expected X-ray flux of a typical luminous Type 1 quasar. To understand the statistical significance of $\Delta \alpha_{\mathrm{OX}}$, we provide the $\sigma$-offset in units $\alpha_{\mathrm{OX}}$ rms scatter from Table 5 of \citet{2006AJ....131.2826S}.

\subsection{\textit{Zwicky Transient Facility} (ZTF) Optical Light Curve Construction}

We obtained the \textit{g/r/i} band light curves for each WLQ using the ZTF forced-photometry service (ZFPS) \citep{2023arXiv230516279M}. {The available ZFPS data provide optical coverage from 2018-03-17 to 2025-07-18. We remove low-quality data by requiring the \texttt{procstatus} flag in the ZFPS light curve output to be 0. To extract calibrated ZTF \textit{g/r/i} band magnitudes for each WLQ in our sample, we compute the source flux from the reference image, apply zero points, and combine the individual epoch difference-image flux and the baseline reference-image flux to estimate the total flux at each epoch. We adopt a signal-to-noise threshold of 3 to define upper-limit photometric data points. We convert the estimated total flux per epoch to magnitudes in the \textit{g}-, \textit{r}- and \textit{i}-bands.} The constructed ZFPS \textit{g/r/i} band light curves can be found in Appendix~\ref{sec:appendixB}.

\section{Variability Summary of WLQ Sample}
Among the 10 WLQs in our sample, two objects exhibited extreme  X-ray variability events, defined in this work as $> 3\sigma_{\mathrm{MAD}}$\footnote{Median absolute deviation (MAD); $\sigma_{\mathrm{MAD}} = 1.483 \times \mathrm{MAD}$. See the discussion in Section 4.1.1.} of the distribution of the ratio between fluxes at two epochs in the \citet{2020MNRAS.498.4033T} radio-quiet quasar sample (see {Section 4.1.1} below). {This extreme X-ray variability threshold corresponds to a full-band X-ray flux factor change of 3.44.} The WLQs with large-amplitude X-ray variations are the previously variable SDSS J1539+3954, and a previously X-ray weak WLQ, SDSS J082508.75+115536.3 (hereafter J0825+1155). J1539+3954 underwent a soft X-ray flux rise of a factor of $\gtrsim 6$ between 2023 August 01 and 2024 October 07 (a factor of $\gtrsim 21$ rise from 2013 December 13), marking the first time we have observed recurrent large-amplitude X-ray variations in a single WLQ. Additionally, J0825+1155 experienced a significant X-ray flux rise by a factor of $\gtrsim 14$ between 2019 September 28 and 2023 January 21. The other WLQs in our sample did not exhibit extreme X-ray variability at the $> 3\sigma_{\mathrm{MAD}}$ level throughout our systematic monitoring campaign. Thus, with the observed large-amplitude variations of J0825+1155, the fraction of WLQs in our sample with observed extreme X-ray variability increases to 2 out of 10. 

{To estimate the true population fraction of WLQs exhibiting extreme X-ray variability from our sample of 10 WLQs, we calculate Bayesian binomial confidence intervals as outlined in \citet{2011PASA...28..128C}. \citet{2011PASA...28..128C} recommend the use of Bayesian confidence intervals on binomial population proportions estimated from the observed population fraction, $\hat{p}$, in small-to-intermediate sized samples. This method utilizes the quantiles of the beta distribution, as derived from the normalized likelihood of observing $\hat{p}$ for a given value of $p$, the true population proportion, and a Bayes-Laplace uniform prior, where $P_{\mathrm{prior}}(p) =  1$ over $0 < p < 1$ assuming all values of $p$ are equally probable, as we have no prior constraints on the true population proportion.}  From the observed population fraction of 0.20, we find the 68\% confidence interval to be 12.8\text{--}37.2\%, or $0.20^{+0.17}_{-0.07}$. 

\begin{figure}[ht!]
\includegraphics[width=0.45\textwidth]{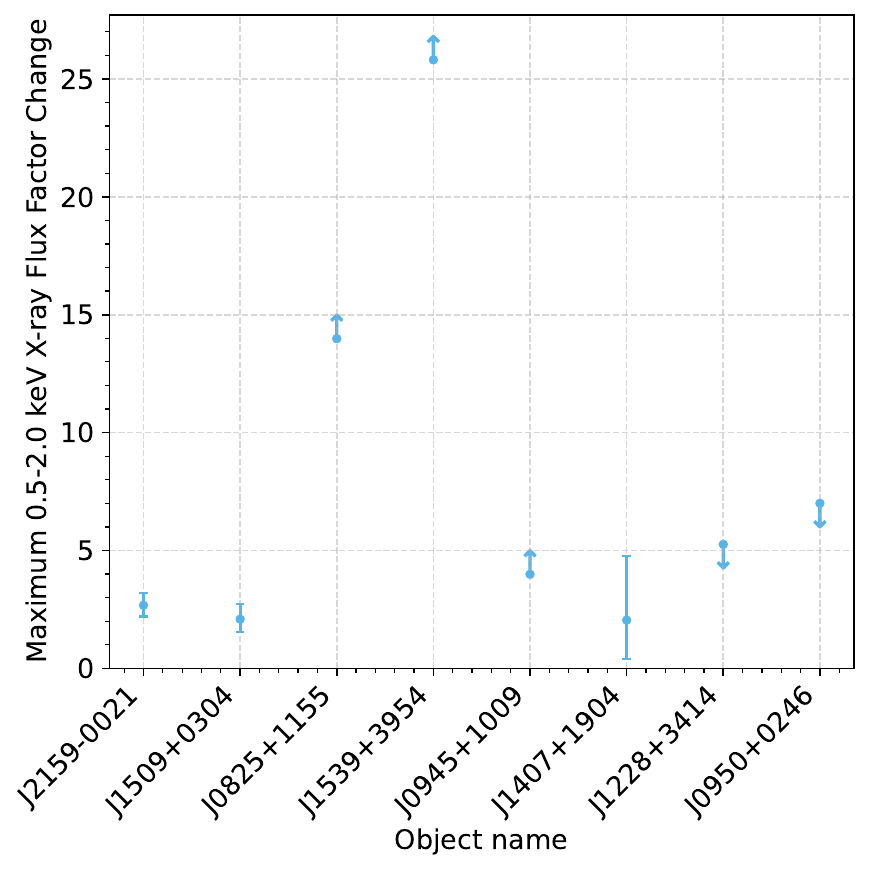}
\caption{Maximum 0.5--2.0 keV X-ray flux-factor change. We are unable to constrain the maximum flux-factor change for two of ten WLQs in our sample (SDSS J110409.96+434507.0 and SDSS J163810.07+115103.9)} when the flux ratios are calculated from two X-ray limits.
\label{fig:xray_flux_factors}
\end{figure}

A summary of the maximum soft X-ray flux-factor change is displayed in Figure~\ref{fig:xray_flux_factors} for each quasar where a flux-factor change can be constrained. For two WLQs in the sample, SDSS J110409.96+434507.0 and SDSS J163810.07+115103.9, we cannot constrain the maximum flux-factor change, as the flux ratios are calculated between X-ray limits. However, we can conclude that these WLQs did not experience a flux rise of a factor greater than 3–8 (relative to their earlier upper-limit measurements) over the duration of this monitoring campaign. This constraint is set by the sensitivity achieved by the exposure times of the campaign. 
  
Within the context of the TDO model, we might expect to observe a transition between the X-ray weak and X-ray normal states to be accompanied by a softening in spectral shape due to reduced X-ray absorption and Compton reflection (see Section 1.1). When able to provide well-constrained measurements for the effective power-law photon index ($\left<1 \sigma \right> < 0.5$), we find the spectral shape to be consistent with the TDO model. As shown in a detailed spectral analysis of the first extreme variability event of J1539+3954 in 2019, the spectral shape is $\Gamma_{\mathrm{eff}} = 2.0 \pm 0.4$ when the WLQ is in an X-ray normal state, with $\Delta \alpha_{\mathrm{OX}} = 0.11$ \citep{2020ApJ...889L..37N}. We find, with a larger uncertainty, that the second transition of J1539+3954 from an X-ray weak to X-ray normal state is associated with a $\Gamma_{\mathrm{eff}} = 2.0 \pm 0.6$, and $\Delta \alpha_{\mathrm{OX}} = 0.06$. While the large uncertainties on measured $\Gamma_{\mathrm{eff}}$ values for J0825+1155 ($\Gamma_{\mathrm{eff}} = 2.8^{+0.7}_{-1.0}$ and $\Delta \alpha_{\mathrm{OX}} = 0.04$ coincident with the observed large-amplitude X-ray variation in 2023) limit our ability to draw conclusions about the nature of the spectral change associated with the observed extreme X-ray variability event of this WLQ, we note that the $\Gamma_{\mathrm{eff}}$ value is consistent with the rapid accretion expected for WLQs. Improved spectral information is needed to definitively state if the X-ray spectral variability matches the expected behavior suggested by the TDO model. 

We investigate if the extremely X-ray variable WLQs J0825+1155 and J1539+3954 stand out in the $\mathrm{C}\,\text{\scriptsize IV}$ parameter space by calculating the $\mathrm{C}\,\text{\scriptsize IV}\ ||$ Distance parameter for each WLQ in the sample. We utilize values of rest-frame $\mathrm{C}\,\text{\scriptsize IV}$ equivalent width and $\mathrm{C}\,\text{\scriptsize IV}$ blueshift from \citet{2018MNRAS.480.5184N}, \citet{2015ApJ...805..122L}, and \citet{2012ApJ...747...10W}, and follow the $\mathrm{C}\,\text{\scriptsize IV}\ ||$ Distance calculation outlined by \citet{McCaffreyTrevor2021CD} to derive $\mathrm{C}\,\text{\scriptsize IV}\ ||$ Distance for each WLQ. We find no apparent connection between the $\mathrm{C}\,\text{\scriptsize IV}\ ||$ Distance and the maximum soft X-ray flux-factor change, or the amplitude of X-ray variability.

We present contemporaneous optical light curves from the Zwicky Transient Facility in the $g/r/i$ bands in Appendix~\ref{sec:appendixB}, probing the photometric properties of the rest-frame UV emission of each WLQ. {To quantify the optical variability of the WLQ sample in a basic manner, we compute the fractional variability amplitude, $F_{\mathrm{var}}$, from the $g$- and $r$-band ZTF light curves of each WLQ.\footnote{{The fractional variability amplitude is given by $F_{\mathrm{var}}=~\frac{1}{\bar{x}}\sqrt{S^2- \sigma_{\mathrm{err}}^2}$, where $S^2$ is the total variance of the light curve, $\sigma_{\mathrm{err}}^2$ is the mean squared measurement error, and $\bar{x}$ is the mean flux of the light curve \citep{2002ApJ...568..610E,2003MNRAS.345.1271V}.}} We exclude the ZTF $i$-band photometric data from this analysis due to the sparse $i$-band coverage across the WLQ sample. We find median $F_{\mathrm{var}}$ values of $0.055 \pm 0.005$ and $0.038 \pm 0.003$ for the $g$- and $r$-bands, respectively.\footnote{{$1\sigma$ errors on the median $F_{\mathrm{var}}$ values are estimated using bootstrap resampling.}}}

{The fractional variability amplitudes of the extremely X-ray variable WLQs J1539+3954 and J0825+1155 are $0.061 \pm 0.001$ and $0.028 \pm 0.002$ in the $g$-band, respectively, and $0.038 \pm 0.001$ and $0.036 \pm 0.001$ in the $r$-band, respectively. The $g$- and $r$-band variability of J1539+3954 is consistent with the WLQ sample medians, indicating a lack of enhanced optical/UV variability despite the strong X-ray variability, while J0825+1155 falls below the average fractional variability amplitude of the total WLQ sample. The presence of strong X-ray variability and average-to-weak optical variability from J1539+3954 and J0825+1155 supports the TDO model, as the X-ray shielding material is on small scales and would not drive the optical variability originating on larger scales.}

{Previous work has shown that WLQs display milder optical variability than a population of typical quasars \citep{2025MNRAS.538L..83K}. To compare the fractional variability amplitudes of our WLQ sample to a population of typical radio-quiet quasars, we created a comparison sample of 50 non-BAL, radio-quiet quasars from the \citet{2011ApJS..194...45S} SDSS quasar catalog. For each WLQ in our sample, we selected the 50 closest objects in redshift from the \citet{2011ApJS..194...45S} catalog. From this subset, we then selected the 5 objects most closely matched in $i$-band absolute magnitude, $M_i$, to the WLQ and added them to the optical variability comparison sample. We confirm that the resulting comparison sample is well matched to our WLQ sample in both redshift and luminosity by performing Kolmogorov-Smirnov (K-S) tests on the redshift and luminosity distributions of the two samples. We constructed the ZTF optical light curves of the comparison sample following the procedure outlined in Section 3.3 to calculate the median fractional variability amplitude in the $g$- and $r$-band. We find median $F_{\mathrm{var}}$ values of $0.062 \pm 0.007$ and $0.050 \pm 0.004$ for the $g$- and $r$-bands of the comparison sample, respectively. Because the measured $F_{\mathrm{var}}$ values for the typical quasar comparison sample are higher than those reported for the WLQ sample, we conclude that WLQs exhibit milder optical variability than typical radio-quiet quasars, consistent with the findings of \citet{2025MNRAS.538L..83K}.}

Both J1539+3954 and J0825+1155 went into Sun block and were unobservable from the ground during their observed X-ray variability. We were thus unable to secure rest-frame UV spectroscopic and photometric support from, e.g., the Hobby-Eberly Telescope (HET) and ZTF during the time period of interest. 

\subsection{Comparing the Frequency of Extreme X-ray Variability in WLQs to Radio-Quiet Quasars}

\subsubsection{Basic Methodology}
To robustly assess if WLQs as a population exhibit extreme X-ray variability more frequently than the general population of Type 1 radio-quiet quasars, we compare our sample of WLQs to the sample of typical Type 1 blue quasars from \citet{2020MNRAS.498.4033T}. The full sample from \citet{2020MNRAS.498.4033T} is composed of quasars from SDSS DR14Q \citep{2018A&A...613A..51P} and spectroscopically confirmed quasars from \citet{2015ApJS..219...39R} that have \textit{Chandra} X-ray counterparts. \citet{2020MNRAS.498.4033T} required their sample to be radio-quiet ($R \leq 30$) and to exclude BALs. 

\citet{2020MNRAS.498.4033T} utilize the 0.5--8.0 keV full-band count flux (in $\mathrm{cts \ cm^{-2} \ s^{-1}}$) to trace changes in the X-ray emission of their sample, as opposed to physical flux units (in $\mathrm{ergs \ cm^{-2} \  s^{-1}}$) in order to avoid introducing uncertainties from fitting the X-ray spectra. To make a one-to-one comparison to the \citet{2020MNRAS.498.4033T} quasar sample, we derive the full-band count flux for all \textit{Chandra} observations of our WLQ sample. We used the CIAO tool FLUXIMAGE to create full-band exposure maps (in units of $\mathrm{photons \ cm^2 \ s}$) for each observation. We then calculated the full-band count flux from the full-band aperture-corrected net counts and the average exposure-map pixel value in the 2\arcsec \ source region extracted from the exposure maps. 

\begin{figure}[ht!]
\centering

\includegraphics[width=0.45\textwidth]
{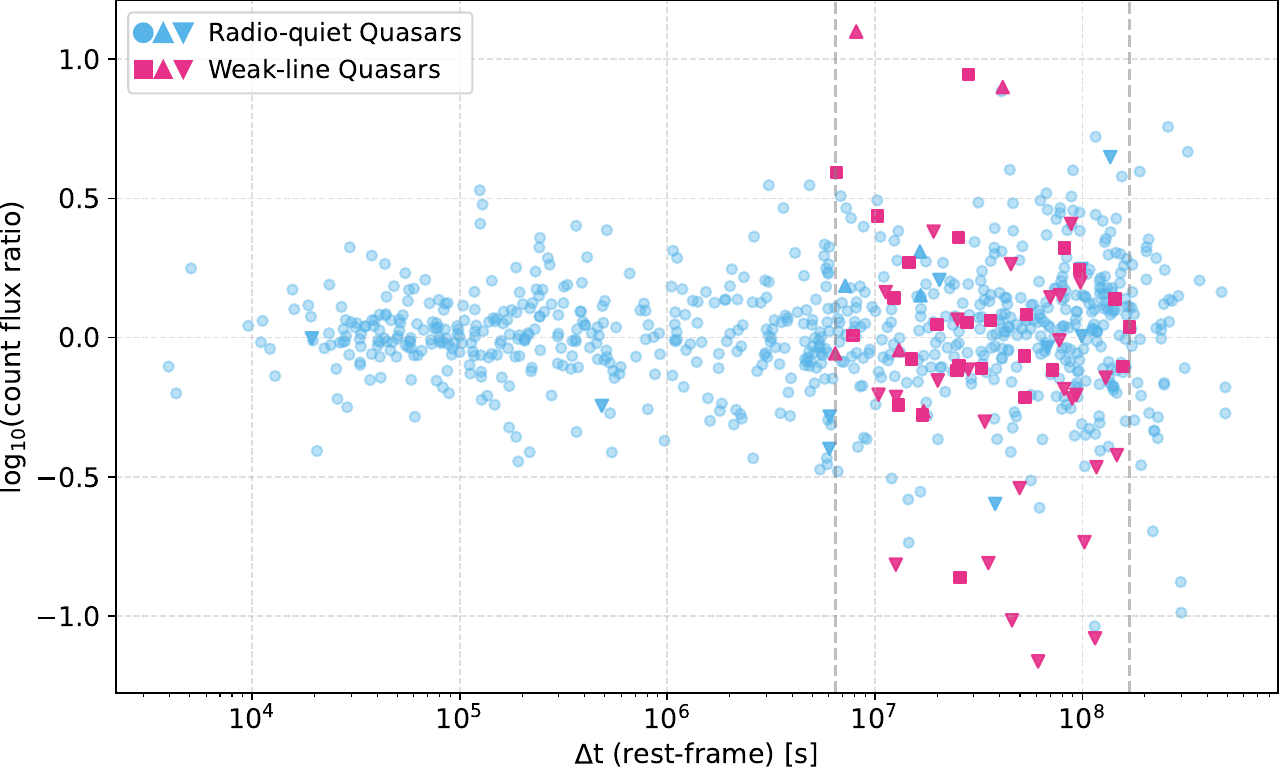}
\caption{Logarithm of the full-band count flux ratio between two observations as a function of the rest-frame time difference between the start of the observations. The down-sampled population of radio-quiet quasars from \citet{2020MNRAS.498.4033T} is shown in blue, with circles as detected values. Our sample of WLQs is shown in pink, with detected values marked as squares. The upper and lower limits of both samples are represented with upward (downward) pointing triangles as upper (lower) limits. The \citet{2020MNRAS.498.4033T} sample contains 556 observation pairs across 216 quasars, and the WLQ sample contains 59 observation pairs of 8 quasars. We note that 2 of the WLQs in the sample were undetected across all epochs, and as such we are unable to constrain the amount of variability between epochs for those objects. The dotted grey lines indicate the minimum and maximum values of $\Delta t$ of the chosen time bin to compare the populations. $\Delta t_{\mathrm{min}} = 6.44 \ \mathrm{Ms}$ and $\Delta t_{\mathrm{max}} = 169.11 \ \mathrm{Ms}$. }
\label{fig:scatter_plot}
\end{figure}

As a measure of the amplitude of X-ray variability, we calculate the ratio of full-band count flux between each unique pair of observations per quasar. We quantify the timescale of the variability by calculating the difference between the rest-frame start times of the two epochs for each observation pair. For each object, there are $N(N-1)/2$ permutations of unique observation pairs, where $N$ is the total number of epochs for a given quasar. When one observation is an X-ray limit, we provide the 90\% confidence level limit on the derived count flux ratio. If the X-ray limit is in the numerator (denominator) of the ratio, there is an upper (lower) limit on the count flux ratio. If both observations are X-ray limits, no constraints can be placed on the count flux ratio between those two epochs. {We present each permutation of observation pairs in the WLQ sample and their detection status (unconstrained, upper limit, lower limit, or detection) in Appendix~\ref{sec:appendixC}}. To ensure no single object has an excessive effect on the overall sample, we use the down-sampled set of observation pairs from \citet{2020MNRAS.498.4033T}, where objects with $N > 5$ observations are down-sampled to include just 5 epochs in the distribution of count flux ratios. The distribution of $\mathrm{log}_{10}(\mathrm{count \ flux \ ratio})$ as a function of the rest-frame timescale probed by the observation pairs for the \citet{2020MNRAS.498.4033T} sample of radio-quiet quasars and the sample of WLQs is displayed in Figure~\ref{fig:scatter_plot}.

\begin{figure}[ht!]
    \centering
    \includegraphics[width=\linewidth]{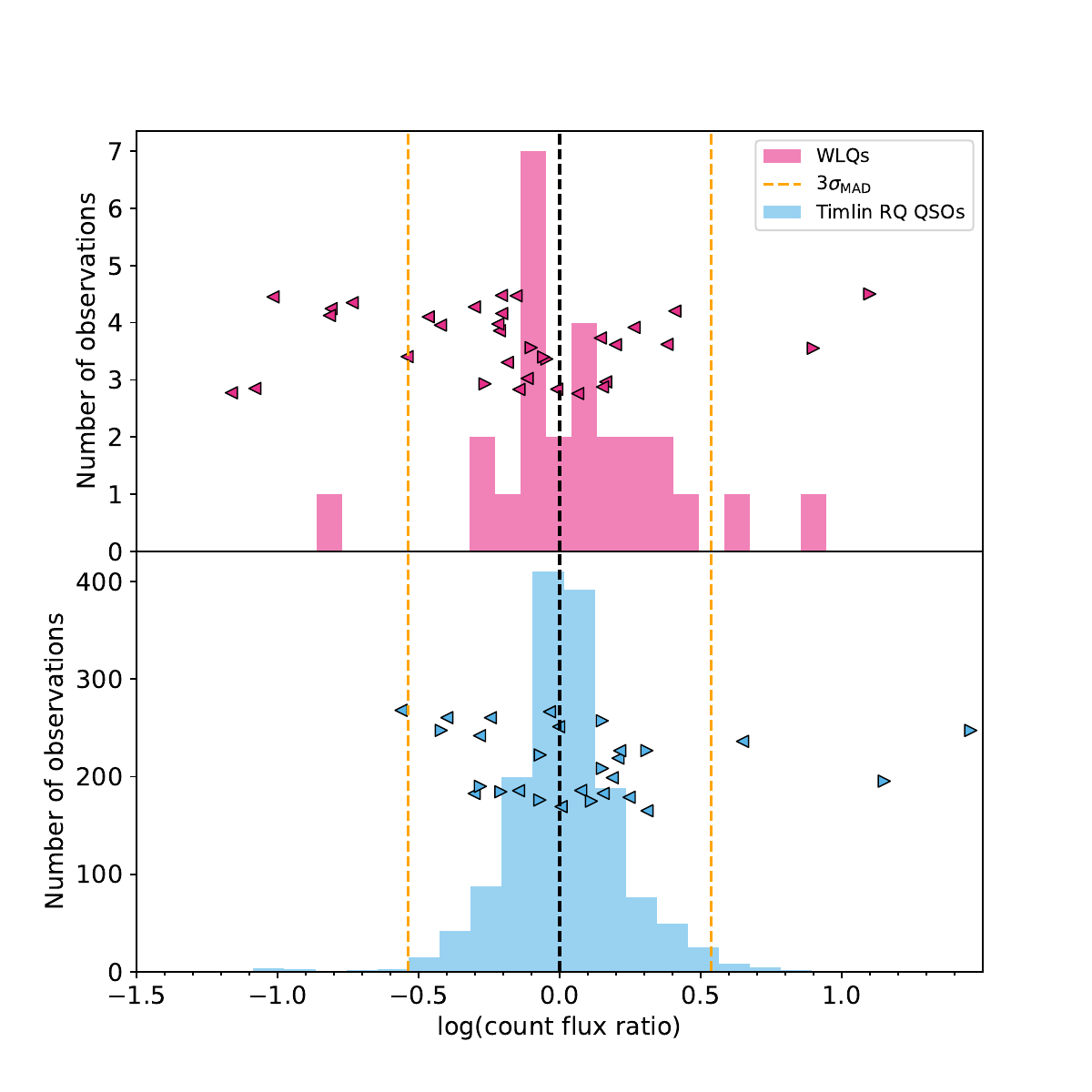}
\caption{Distribution of full-band count flux ratios for observations of time separations $6.44 \ \mathrm{Ms} \leq \Delta t \leq 169.11 \ \mathrm{Ms}$. {The left-pointing (right-pointing) arrows represent the upper (lower) X-ray limits, and are analogous to the X-ray limits presented in Figure~\ref{fig:scatter_plot}.} The $3\sigma_{\mathrm{MAD}} = 0.536$ threshold is denoted by the dashed orange lines. }
\label{fig:log_hist}
\end{figure}

To compare the amplitude of variability on comparable timescales between the two populations, we limit our comparison to the observation pairs from the \citet{2020MNRAS.498.4033T} sample that fall within the minimum and maximum $\Delta t$ rest-frame time difference of the WLQ sample ($6.44 \  \mathrm{Ms} \leq \Delta t \leq 169.11 \ \mathrm{Ms}$). The distributions of $\mathrm{log}_{10}(\mathrm{count \ flux \ ratio})$ for this range of time-scales are displayed in Figure~\ref{fig:log_hist}. We find that the $\Delta t$ distributions of the WLQ and \citet{2020MNRAS.498.4033T} samples are not materially different following from a K-S test upon the two samples, indicating that the timescales are similarly distributed within the considered $\Delta t$ range.

\citet{2020MNRAS.498.4033T} showed that the distributions of count flux ratios for their radio-quiet quasar sample deviated from a Gaussian distribution in the wings, and thus opted to use the median absolute deviation (MAD) as a robust statistical measure of the spread of the distribution. The MAD statistic is more resistant to outliers than the standard deviation of the sample and is appropriate for non-Gaussian distributions, e.g., \citet{10.1002/0470010940}. The MAD is related to the Gaussian standard deviation by $\sigma_{\mathrm{MAD}} = 1.483 \times \mathrm{MAD}$. In this work, we define count flux ratios of $ \geq 3\sigma_{\mathrm{MAD}}$ as extremely variable measurements, where $3\sigma_{\mathrm{MAD}} = 0.536$ in the \citet{2020MNRAS.498.4033T} sample. In the WLQ sample, 12 out of 59 observation pairs contain count flux ratios $ \geq 3\sigma_{\mathrm{MAD}}$ and are classified as extremely X-ray variable.

\subsubsection{Fisher's Exact Test}

We employ Fisher’s exact test \citep{10.2307/2340521} to determine the probability of observing 12 or more $3\sigma_{\mathrm{MAD}}$ outliers in our WLQ sample if we take the underlying distribution to be consistent with the \citet{2020MNRAS.498.4033T} sample of radio-quiet quasars. Because Fisher’s exact test intrinsically accounts for uncertainty in the observed count flux ratios by providing the probability of observing data that are as extreme as or more extreme than those observed, we use the reported X-ray limits directly in this test. In Table~\ref{tab:contingency}, we organize the data into a $2 \times 2$ contingency table that lists, for both the WLQ and radio-quiet quasar samples, the number of observation pairs classified as $3\sigma_{\mathrm{MAD}}$ outliers and the number of non-outlier observation pairs. We performed a one-tailed Fisher’s exact test to derive the probability of observing data as extreme (or more extreme) as that presented in Table~\ref{tab:contingency}, under the {null hypothesis} that both samples are equally likely to exhibit extreme variability behavior. {The one-sided $p$-value gives the probability of observing $x \geq a$ outliers, where $a$ is the number of observed $3\sigma_{\mathrm{MAD}}$ outliers in the WLQ sample.} We find a $p$-value of $1.02 \times 10^{-5}$, indicating that the WLQ sample has a different underlying distribution of variability amplitudes than the \citet{2020MNRAS.498.4033T} sample that results in a higher frequency of observed $3\sigma_{\mathrm{MAD}}$ outliers. 

We additionally calculate the odds ratio of the contingency table, defined as the ratio of the odds of observing a $3\sigma_{\mathrm{MAD}}$ outlier in the WLQ sample to that in the \citet{2020MNRAS.498.4033T} sample. {The odds ratio allows us to probe if the WLQ sample exhibits extreme X-ray variability events more frequently than the typical radio-quiet quasar sample, where an odds ratio greater than 1.0 indicates that the probability of observing extreme X-ray variability in the WLQ sample is higher than in the quasar sample from \citet{2020MNRAS.498.4033T}.} We find an odds ratio of 6.8, indicating that the odds of observing an extreme X-ray variability event in WLQs is 6.8 times {higher than} that of the \citet{2020MNRAS.498.4033T} radio-quiet quasar sample. The 95\% confidence interval on the odds ratio is (2.85, 15.70), falling entirely above an odds ratio of 1.0, thus supporting the conclusion that WLQs more frequently experience extreme X-ray variability events than the general population of typical radio-quiet quasars; furthermore, the odds ratio {of 6.8} provides a basic quantification of the magnitude of the effect for the first time. 

We test whether this result is sensitive to the choice of the $3\sigma_{\mathrm{MAD}}$ threshold defining extreme X-ray variability amplitudes by conducting Fisher’s exact tests for a range of thresholds from $2\sigma_{\mathrm{MAD}}\text{--}4\sigma_{\mathrm{MAD}}$. We find that the odds ratio (and its 95\% confidence interval) is greater than 1.0 for all given thresholds, ranging between 2.73 ($2\sigma_{\mathrm{MAD}}$) and 15.84 ($4\sigma_{\mathrm{MAD}}$).

\begin{table}[ht!]
\centering
\caption{Fisher's Exact Test Contingency Table \label{tab:contingency}}
\begin{tabular}{lccc}
\hline\hline
 & WLQs & Timlin QSOs & Row Total \\
\hline
$3\sigma_{\mathrm{MAD}}$ outliers & 12 & 20  & 32  \\
Non-outliers & 47 & 536 & 583 \\
\hline
Column total  & 59 & 556 & 615 \\
\hline\hline
\end{tabular}

\tablecomments{\noindent
Categorical $2 \times 2$ contingency table displaying the number of observation pairs
classified as outliers and non-outliers, defined by a $3\sigma_{\mathrm{MAD}}$ threshold,
for both the WLQ and \citet{2020MNRAS.498.4033T} radio-quiet quasar samples. {The 12 outlier observation pairs in the WLQ sample come from J1539+3954 (8 outliers), J0825+1155 (3 outliers), and SDSS J140701.59+190417.9 (1 outlier).}
}
\end{table}

\subsubsection{Censored Two-Sample Tests}
For a statistical treatment of the entire distribution of variability amplitudes between samples, instead of only considering the outliers, we need to apply a univariate two-sample test. Following from the results of Fisher's exact test in Section 4.1.2, we would expect the overall distribution of the WLQ sample to more frequently sample the extreme wings of the distribution. While the K-S test and the Anderson-Darling two-sample test \citep{10.2307/2280095,10.1214/aoms/1177729437} are unable to account for censored data, there exist univariate singly-censored two-sample tests like the logrank test and the Peto-Peto test \citep{2012msma.book.....F}. However, in both samples, the distributions of $|\mathrm{log}_{10}(\mathrm{count \ flux \ ratio})|$ contain both upper and lower limits, making the distributions doubly-censored. 

Since we cannot straightforwardly employ the singly-censored two-sample tests on the doubly-censored distributions, we must first convert our samples into a singly-censored problem by drawing random values from a uniform distribution to replace the upper-limit values in each sample. We randomly sample and replace the upper-limit values, as opposed to the lower-limit values, because the upper limits make up a small percentage of the combined data samples ($\approx20\%$ of the WLQ sample corresponding to 12/59 observation pairs, and $\approx2\%$ of the radio-quiet quasar sample corresponding to 14/556 observation pairs). In addition, these upper-limit values largely fall within the observed distributions of their respective samples, allowing us to randomly sample values for the upper limits from a well-defined uniform distribution dictated by the two samples. The average value of the 90\% confidence upper limits, $ \left\langle \mathrm{UL_{WLQ}} \right\rangle = 0.187$ and $\left\langle \mathrm{UL_{RQ \  QSO}} \right\rangle = 0.227$, fall within $3\sigma_{\mathrm{MAD}} = 0.536$ of the \citet{2020MNRAS.498.4033T} distribution.  The uniform distribution from which we drew values for the upper limits was defined to be between the minimum value of $|\mathrm{log}_{10}(\mathrm{count \ flux \ ratio})|$ (of the sample distribution containing the given upper limit) and the 90\% confidence upper-limit value. The resulting samples are singly-censored, containing lower-limit, detected, and resampled upper-limit values. Using this random-sampling method for the upper-limit values, we created 5,000 instances of  $|\mathrm{log}_{10}(\mathrm{count \ flux \ ratio})|$ distributions to conduct censored two-sample tests between the WLQs and the radio-quiet quasars of \citet{2020MNRAS.498.4033T}. 

We conduct all censored two-sample tests using the Astronomy SURVival analysis (ASURV)\footnote{\url{https://github.com/rsnemmen/asurv}} \citep{1985ApJ...293..192F,1986ApJ...306..490I, 1990BAAS...22..917I} package. \citet{2012msma.book.....F} suggest the application of multiple censored two-sample tests with different weightings of censored values in the event that the underlying population distributions are unknown. Here, we utilize the logrank test, the Peto-Peto test, and the Peto-Prentice test \citep{2012msma.book.....F}. We compute the median $p$-value, representing the probability that the two samples belong to the same underlying population distribution, for each two-sample test across the 5,000 pairs of WLQ and \citet{2020MNRAS.498.4033T} radio-quiet quasar samples. We find that the probability of the WLQ and radio-quiet quasar samples belonging to the same underlying distribution is $\lesssim 10^{-4}\text{--}10^{-3}$ for the range of applied two-sample tests. We thus conclude that the distribution of variability amplitudes of WLQs and the general population of radio-quiet quasars are statistically different to a high significance.

\section{Conclusions \& Future Work}

\subsection{Summary of Results}
In this work, we have presented the X-ray photometric analyses of 10 WLQs as part of a systematic monitoring campaign for extreme X-ray variability events in WLQs. We performed comparative statistical analyses of the WLQ and \citet{2020MNRAS.498.4033T} radio-quiet quasar samples. The key results from this work are summarized below: 

\begin{enumerate}

 \item We selected 10 WLQs with $\mathrm{C}\,\text{\scriptsize IV}$ REW $\lesssim 10$ \AA \ from the representative WLQ sample of \citet{2018MNRAS.480.5184N} to observe over \textit{Chandra} Cycles \mbox{23--25}. We also utilized 1--3 archival \textit{Chandra} epochs per WLQ to assess the long-term X-ray variability of the sample, resulting in 4--6 total epochs per WLQ (see Section 2).

 \item We performed aperture photometry to estimate the X-ray flux (in the full, soft, and hard bands) for each observation. We presented the 90\% confidence upper limit in the event that the source was not significantly detected (binomial no-source probability, $P_B > 0.01$). We assess the X-ray weakness of a given observation through the derived $\Delta \alpha_{\mathrm{OX}}$ values (see Section 3).

 \item For the historically variable WLQ in our sample, J1539+3954, we observed recurrent extreme X-ray variability (variability amplitude $> 3\sigma_{\mathrm{MAD}}$ of the distribution of the ratio between fluxes of two epochs in the \citet{2020MNRAS.498.4033T} radio-quiet quasar sample). We observed an X-ray flux factor rise of $\gtrsim 6$ between August 2023  and October 2024, and on a longer timescale, a factor of $\gtrsim 21$ rise between 2013 and 2024. We find an effective power-law photon index of $\Gamma_{\mathrm{eff}} = 2.0 \pm 0.6$ for the 2024 observation of J1539+3954 when transitioning to an X-ray normal state, consistent with a softening in spectral shape due to reduced X-ray absorption and/or Compton reflection. However, further spectral information is required to assess the validity of the TDO model robustly (see Section 4).

 \item We observed the first extreme X-ray variability event (X-ray flux factor rise of $\gtrsim 14$ between September 2019 and January 2023) of the WLQ J0825+1155, which was previously X-ray weak. This observation increased the fraction of the WLQ sample in which we detect evidence of extreme X-ray variability to $0.20^{+0.17}_{-0.07}$, or 12.8\text{--}37.2\%  (see Section 4). 

 \item We conducted a statistical comparison between the X-ray variability amplitudes of WLQs and typical Type 1 blue quasars from \citet{2020MNRAS.498.4033T}. Using Fisher’s exact test, we find that the WLQ sample more frequently produces extreme X-ray variability events to a high significance, with a $p$-value of $1.02 \times 10^{-5}$ and an odds ratio of 6.8. (see Sections 4.1.1, 4.1.2). 

 \item We formulate our data into a singly-censored distribution, randomly sampling upper-limit values of $|\mathrm{log}_{10}(\mathrm{count \ flux \ ratio})|$ from a uniform distribution in order to apply singly-censored two-sample tests. We found the samples of WLQs and \citet{2020MNRAS.498.4033T} radio-quiet quasars to be statistically different at high significance (see Section 4.1.3).
\end{enumerate}

The additional sensitive multi-epoch observations obtained from this \textit{Chandra} monitoring campaign have revealed an increased fraction of WLQs (12.8\text{--}37.2\%) with extreme X-ray variability. Although the source statistics remain limited, the substantial fraction of extremely X-ray variable WLQs, in the context of the TDO model, may suggest that the variability is more likely driven by intrinsic motion of the TDO wind rather than by changes in the height of the TDO disk (see Section 1.1). Additionally, we observed a WLQ to display repeated large-amplitude X-ray variations for the first time in J1539+3954. Under the TDO model, this recurrent variability implies that the physical mechanism driving the changes in the TDO wind can be episodic in nature. While we cannot yet constrain the origin of repeated extreme X-ray variability events in WLQs, this work suggests that the underlying cause of such variability is not due to a stand-alone transient event.

\subsection{Future Work}
An extension of the systematic X-ray monitoring campaign to the other luminous, optically bright, $\mathrm{C}\,\text{\scriptsize IV}$ REW $\lesssim 10$ \AA \ WLQs from the \citet{2018MNRAS.480.5184N} representative sample of WLQs could help further constrain the fraction of WLQs that undergo extreme X-ray variability events and the frequency at which they occur. Future X-ray missions such as NewAthena \citep{2025NatAs...9...36C}, with the order of magnitude improvement in photon-collection capability, will help further constrain the amplitude of extreme X-ray variations. Observations with NewAthena will also allow further assessment of the validity of the TDO model by placing detailed constraints on the change in X-ray spectral shape associated with transitions between the X-ray weak and X-ray normal states in WLQs. The high-resolution spectroscopic capabilities of NewAthena will enable more rigorous testing of the TDO model by revealing the fundamental properties (e.g., velocity of outflows, structure) of WLQs. 

{The TDO model described in this work has built upon the shielding-gas model \citep{2011ApJ...736...28W,2015ApJ...805..122L}, emphasizing the likely importance of the outflowing wind in addition to the thick inner disk.} Evaluating the ability of other physical accretion models to reproduce the distinctive multiwavelength properties of WLQs will allow a more detailed investigation of the nature of WLQs, building upon the insights from the basic TDO model. {For instance, the analytical slim-disk model \citep{1988ApJ...332..646A} with an associated wind and modern numerical simulations of super-Eddington accretion flows (e.g., \citealt{2024arXiv240816856J}) both contain geometrically thick inner disks and/or outflows that could serve as the shielding material. Additionally, comparisons to} the multi-phase, magnetically-dominated quasar accretion-disk model proposed by \citet{2025OJAp....8E..56H}{, as well as an assessment of its ability to reproduce the distinctive observable properties of WLQs, including the extreme X-ray variability highlighted in this work, will help inform a more detailed understanding of the accretion physics and structure of WLQs.}

\begin{acknowledgments}
We thank J.D. Timlin for helpful discussions. MR and WNB acknowledge the support of Chandra X-ray Center grant GO2-23083X and the Penn State Eberly Endowment. B.L. acknowledges financial support from the National Natural Science Foundation of China grant 12573016. FV acknowledges support from "INAF Ricerca Fondamentale 2023 - Large GO grant."

This research has made use of data obtained from the Chandra Data Archive, {contained in the \textit{Chandra} Data Collection (CDC) 609~\dataset[DOI:10.25574/cdc.609]{https://doi.org/10.25574/cdc.609},} and software provided by the Chandra X-ray Center in the application package CIAO.

The presented optical light curves are based on observations obtained with the Samuel Oschin 48-inch Telescope at the Palomar Observatory as part of the Zwicky
Transient Facility project. ZTF is supported by the National Science Foundation under Grant No. AST-1440341 and a
collaboration including Caltech, IPAC, the Weizmann Institute for Science, the Oskar Klein Center at Stockholm University, the
University of Maryland, the University of Washington, Deutsches Elektronen-Synchrotron and Humboldt University, Los Alamos
National Laboratories, the TANGO Consortium of Taiwan, the University of Wisconsin at Milwaukee, and Lawrence Berkeley
National Laboratories. Operations are conducted by COO, IPAC, and UW.

\end{acknowledgments}

\clearpage
\appendix
\restartappendixnumbering

\section{X-ray and Optical Properties of the WLQ Sample}
\label{sec:appendixA}
We present a table containing the \textit{Chandra} Observations and their derived X-ray and optical properties. We outline the contents of Table~\ref{tab:tableA1} below, and the full table is available online in a machine readable format. 

\begin{deluxetable}{ll}[h!]
\tablecaption{X-ray and Optical Properties of the WLQ Sample\label{tab:tableA1}}
\tablehead{Column number & Description}
\startdata
Column (1) & WLQ object name in J2000 coordinate format. \\
Column (2) & \textit{Chandra} observation ID. \\
Column (3) & Observation start date in YYYY-MM-DD format. \\
Column (4) & Full-band (0.5-8.0 keV) effective exposure time in units of ks. \\
Column (5)--(7) & Aperture-corrected source counts in the full (0.5-- 8.0 keV) band (90\% confidence level \\ & upper limits are provided if the source is not detected); 1$\sigma$ upper and lower errors of the \\ & aperture-corrected source counts in the full band. \\
Column (8)--(10) & Aperture-corrected source counts in the soft (0.5-- 2.0 keV) band (90\% confidence level \\ & upper limits are provided if the source is not detected); 1$\sigma$ upper and lower errors of the \\ & aperture-corrected source counts in the soft band. \\
Column (11)--(13) & Aperture-corrected source counts in the hard (2.0-- 8.0 keV) band (90\% confidence level \\ & upper limits are provided if the source is not detected); 1$\sigma$ upper and lower errors of the \\ & aperture-corrected source counts in the hard band. \\
Column (14)--(16) & Band ratio between the hard-band and the soft-band counts (90\% confidence level \\ & upper limits if the source is only detected in the soft-band); 68$\%$ confidence level upper \\ & and lower errors. \\
Column (17)--(19) & 0.5--8 keV effective power-law photon index; 68$\%$ confidence level upper and lower errors. \\
Column (20) & Observed 0.5-2.0 keV flux corrected for Galactic absorption in units of $10^{-15} \ \mathrm{erg \ cm^{-2} \ s^{-1}}$. \\
Column (21) & Rest-frame 2 keV flux density in units of $10^{-32} \ \mathrm{erg \ cm^{-2} \ s^{-1} \ Hz^{-1}}$. \\
Column (22) & Rest-frame 2500 \AA\ flux density in units of $10^{-27} \ \mathrm{erg \ cm^{-2} \ s^{-1} \ Hz^{-1}}$. \\
Column (23) & Logarithm of the 2500 \AA\ luminosity in units of $\mathrm{erg \ s^{-1} \ Hz^{-1}}$. \\
Column (24) & Measured $\alpha_{\mathrm{OX}}$ values. \\
Column (25) & Difference between the measured $\alpha_{\mathrm{OX}}$ values and the expected $\alpha_{\mathrm{OX}}$ values from the \\ & $\alpha_{\mathrm{OX}}-L_{2500\text{\AA}}$ relation in \citet{2007ApJ...665.1004J}. \\
Column (26) & Statistical significance of $\Delta \alpha_{\mathrm{OX}}$. \\
\enddata
\end{deluxetable}

\clearpage
\section{ZTF Optical Light Curves}
\label{sec:appendixB}
\restartappendixnumbering

\begin{figure*}[htp!]
\centering
\includegraphics[width=0.8\textwidth]{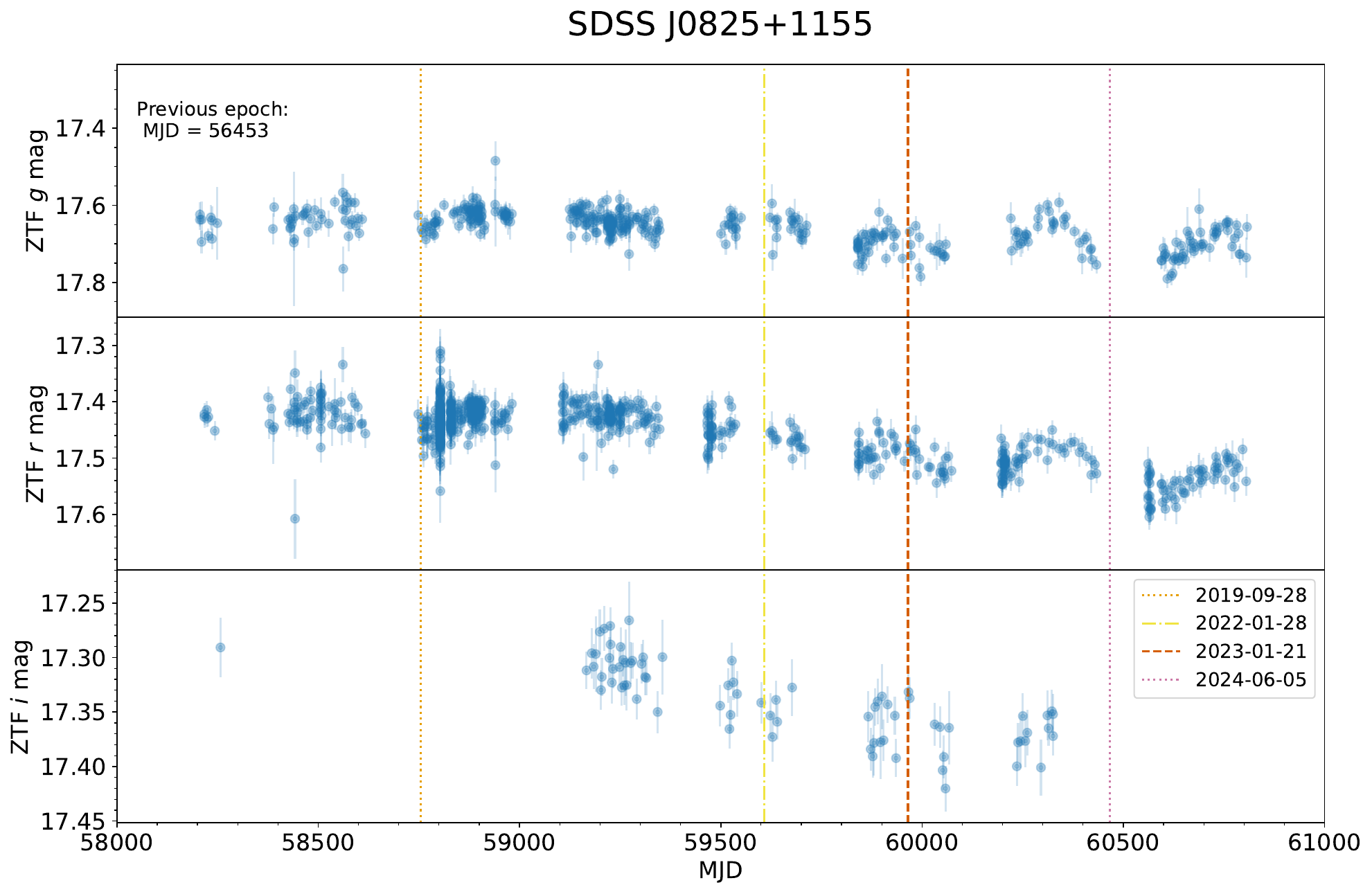}
\caption{ZTF \textit{g}/\textit{r}/\textit{i}-band light curves of the sample of WLQs extracted from ZTF's forced photometry service (ZFPS) \citet{2019PASP..131a8003M}. The vertical lines mark the dates of the \textit{Chandra} observations of a given WLQ. Any \textit{Chandra} epochs that pre-date the ZTF survey are noted in the top left corner of each set of light curves. The complete figure set (10 images) is available in the online journal.}\label{fig:ztf_lcs}
\end{figure*}

{We present the constructed ZTF \textit{g}, \textit{r}, and \textit{i}-band light curves for each WLQ, as described in Section 3.3, in Figure~\ref{fig:ztf_lcs}. The epochs of the \textit{Chandra} X-ray observations are indicated by the vertical lines.}

\FloatBarrier

\noprint{\figsetstart}
\noprint{\figsetnum{B1}}
\noprint{\figsettitle{ZTF \textit{g}/\textit{r}/\textit{i}-band light curves}}

\figsetgrpstart
\figsetgrpnum{B1.1}
\figsetgrptitle{SDSS J0825+1155}
\figsetplot{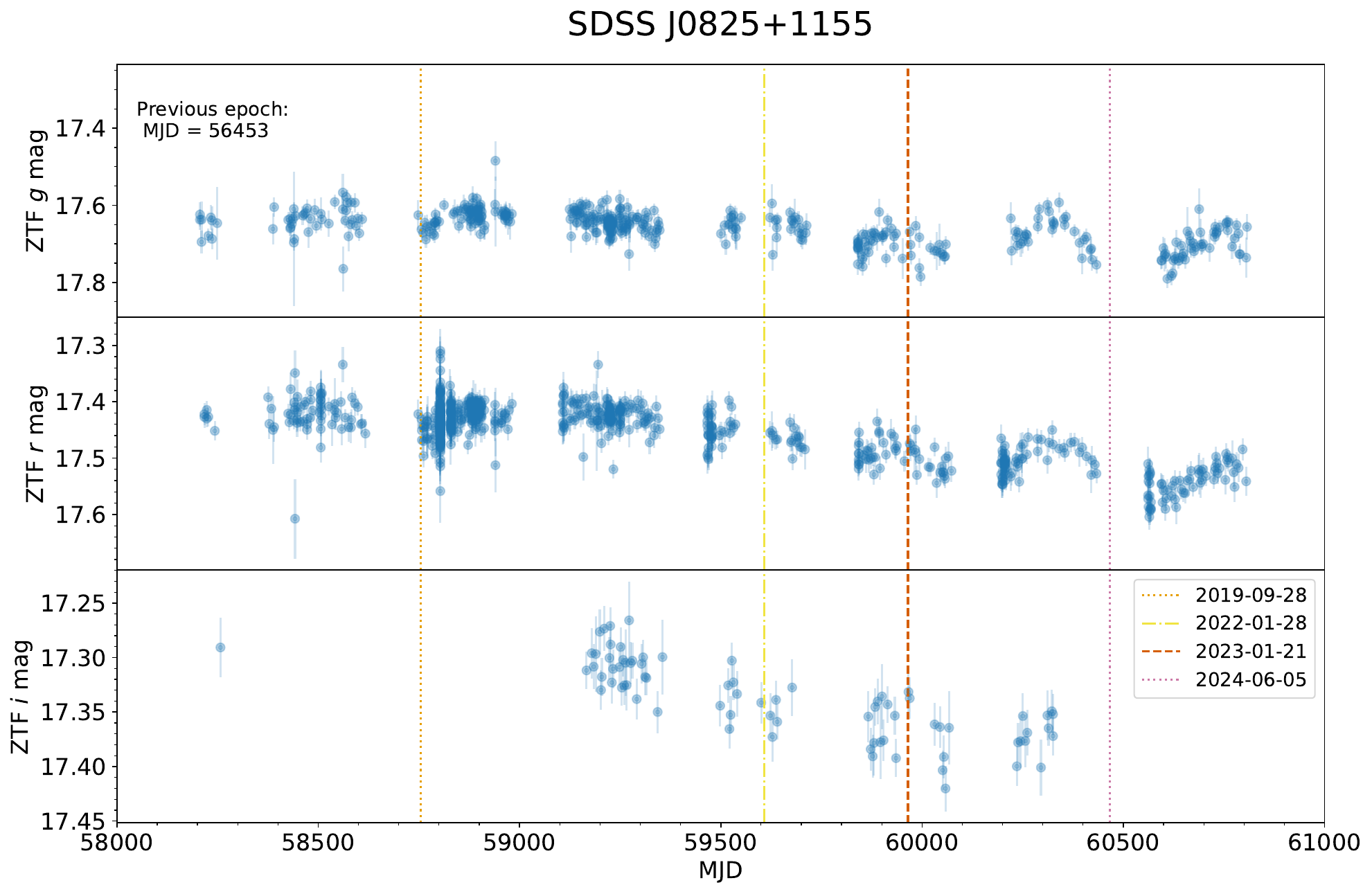}
\figsetgrpnote{ZTF \textit{g}/\textit{r}/\textit{i}-band light curves of SDSS J0825+1155. The vertical lines mark the dates of the \textit{Chandra} observations. Any \textit{Chandra} epochs that pre-date the ZTF survey are noted in the top right corner.}
\figsetgrpend

\figsetend

\clearpage
\section{Permutations of X-ray Flux Ratios in the WLQ Sample}
\label{sec:appendixC}
\restartappendixnumbering

{We present a subset of the permutations of observation pairs in the WLQ sample, as described in Section 4.1.1, in Table~\ref{tab:flux_ratios}. There are $N(N-1)/2$ permutations for each WLQ, where $N$ is the total number of X-ray epochs for a given WLQ. The full set of 93 permutations across the WLQ sample is available in machine-readable format.}

\begin{deluxetable}{cccccc}[h!]
\centering
\tabletypesize{\footnotesize}
\tablecaption{Permutations of WLQ Sample Count Flux Ratios \label{tab:flux_ratios}}
\tablehead{
Object name & 
MJD start time, $i$ & 
MJD start time, $j$ & 
$\Delta t$ & 
log(count flux ratio) &
Detection status\\
(1) & (2) & (3) & (4) & (5) & (6)\\
}
\startdata
SDSS J082508.75+115536.3 & 56453.42 & 58754.82 & 67062583.53 & ... & unconstrained \\
SDSS J082508.75+115536.3 & 56453.42 & 59607.36 & 91905665.72 & ... & unconstrained\\
SDSS J082508.75+115536.3 & 56453.42 & 59965.01 & 102327425.81 & -0.74 & upper \\
\enddata
\tablecomments{(1) WLQ object name in J2000 coordinates. (2) MJD start time of the first ($i$) epoch in the observation pair. (3) MJD start time of the second ($j$) epoch in the observation pair.  (4) Rest-frame start time difference between the two epochs in units of seconds. (5) Logarithm of the full-band count flux ratio between the two epochs. (6) Detection status of the resultant flux ratio. If both epochs are X-ray limits, the detection status of the flux ratio is ‘unconstrained.’ If only one epoch is an X-ray limit, the detection status of the flux ratio is itself an X-ray limit (‘upper’ or ‘lower’). If both epochs are detected, the detection status of the flux ratio is ‘detected.’ Table~\ref{tab:flux_ratios} is published in its entirety in machine-readable format, and a segment is shown here to highlight its format and content.}
\end{deluxetable}

\clearpage
\bibliography{main}{}
\bibliographystyle{aasjournalv7}

\end{document}